\documentclass[journal]{IEEEtran} 
\usepackage{graphicx,latexsym,color,amsfonts}
\usepackage[intlimits]{amsmath}
\usepackage{xcolor}
\usepackage[colorlinks]{hyperref}
\usepackage{tikz}
\usetikzlibrary{shapes}
\usetikzlibrary{plotmarks}
\usepackage[normalem]{ulem}
\usepackage[utf8]{inputenc}
\usepackage[nolist]{acronym}

\providecommand*{\mrm}[1]{\mathrm{#1}}

\DeclareMathAccent{\ring}{\mathalpha}{operators}{"17}

\providecommand*{\unit}[1]{\ensuremath{\mrm{\,#1}}}
\providecommand*{\eu}{\ensuremath{\mrm{e}}}

\providecommand*{\ju}{\ensuremath{\mrm{j}}}

\newcommand{\partder}[2]{\frac{\partial#1}{\partial#2}}

\renewcommand{\vec}[1]{{\boldsymbol#1}}
\newcommand{\mat}[1]{{\mathbf{#1}}}
\newcommand{\R}{\mathbb{R}}

\newcommand{\qtext}[1]{\quad\text{#1}}

\newcommand{\Id}{\mat{1}}							
\newcommand{\eig}{\mathop{\mrm{eig}}}
\newcommand{\svd}{\mathop{\mrm{svd}}}
\providecommand*{\diffS}{\operatorname{dS}\!}

\newcommand{\reg}{\varOmega}
\newcommand{\basv}{\vec{\psi}}
\newcommand{\sphind}{\upsilon}

\newcommand{\rv}{\vec{r}}
\newcommand{\psiv}{\vec{\psi}}

\newcommand{\zvh}{\hat{\vec{z}}}
\newcommand{\rvh}{\hat{\vec{r}}}
\newcommand{\kvh}{\hat{\vec{k}}}
\newcommand{\evh}{\hat{\vec{e}}}
\newcommand{\hvh}{\hat{\vec{h}}}
\newcommand{\Fm}{\mat{F}}
\newcommand{\Jm}{\mat{I}}
\newcommand{\Jmt}{\tilde{\mat{I}}}
\newcommand{\Zm}{\mat{Z}}
\newcommand{\Xm}{\mat{X}}
\newcommand{\Rm}{\mat{R}}
\newcommand{\Om}{\mat{0}}

\newcommand{\Um}{\mat{U}}

\newcommand{\Vm}{\mat{V}}
\newcommand{\Sm}{\mat{S}}

\newcommand{\Sigmam}{\mat{\Sigma}}
\newcommand{\Psim}{\mat{\Psi}}

\newcommand{\tran}{\mrm{T}}

\newcommand{\Xme}{\mat{X}_{\mrm{e}}}
\newcommand{\Xmm}{\mat{X}_{\mrm{m}}}

\newcommand{\herm}{\mrm{H}}

\newcommand{\Rml}{\mat{R}_{\Omega}}
\newcommand{\Rmlf}{\mat{\Upsilon}}
\newcommand{\Rmr}{\mat{R}_{\mrm{r}}}

\newcommand{\Pl}{P_{\mrm{\Omega}}}

\newcommand{\Prad}{P_{\mrm{r}}}

\newcommand{\Aeff}{A_{\mrm{eff}}}
\newcommand{\Across}{A_{\mrm{cross}}}

\newcommand{\ie}{\textit{i.e.}}
\newcommand{\eg}{\textit{e.g.}}
\newcommand{\cf}{\textit{cf.}}

\newcommand{\Rs}{R_\mrm{s}}

\newcommand{\minimize}{\mathop{\mrm{minimize}}}

\newcommand{\maximize}{\mathop{\mrm{maximize}}}

\newcommand{\subto}{\mrm{subject\ to}}

\newcommand{\nul}{\nu_{\mrm{l}}}
\newcommand{\eigv}{\gamma}

\newcommand{\Ohmps}{\unit{\Omega}/\square}
\newcommand{\Rop}{\mathop{\mrm{R}}}
\newcommand{\uvop}{{\mat{u}}}

\newcommand{\Jv}{\vec{J}}

\newcommand{\Aff}{A_\mrm{eff}}
\newcommand{\ZVAC}{Z_0}

\hypersetup{
    colorlinks,
    linkcolor={red!30!black},
    citecolor={blue!40!black},
    urlcolor={blue!20!black}
}

\newcommand{\nuopt}{\nu_{\mrm{opt}}}
\newcommand{\Gmax}{G_{\mrm{ub}}}
\newcommand{\Gmaxr}{G_{\mrm{ub,r}}}

\newacro{MoM}[MoM]{method of moments}
\newacro{MOO}[MOO]{multiobjective optimization}
\newacro{CM}[CM]{characteristic mode}
\newacro{PEC}[PEC]{perfect electric conductor}
\newacro{PMC}[PMC]{perfect magnetic conductor}
\newacro{EP}[EP]{eigenvalue problem}
\newacro{GEP}[GEP]{generalized eigenvalue problem}
\newacro{EFIE}[EFIE]{electric field integral equation}
\newacro{SVD}[SVD]{singular value decomposition}
\newacro{RWG}[RWG]{Rao-Wilton-Glisson}
\newacro{EM}[EM]{electromagnetic}
\newacro{DOF}[DOF]{\mbox{degrees-of-freedom}}
\newacro{QCQP}[QCQP]{quadratically constrained quadratic program}
\newacro{SDP}[SDP]{semidefinite program}
\newacro{FMM}[FMM]{fast multipole method}
\newacro{BoR}[BoR]{body of revolution}

\definecolor{metal}{RGB}{218,165,32}

\begin{document}
\title{Maximum Gain, Effective Area, and Directivity}
\author{Mats~Gustafsson,~\IEEEmembership{Senior Member,~IEEE}
        and
        Miloslav~Capek,~\IEEEmembership{Senior Member,~IEEE}	    
\thanks{Manuscript received  \today; revised \today.
This work was supported by the Swedish Foundation for Strategic Research (SSF) grant no.~AM13-0011. The work of M.~Capek was supported by the Ministry of Education, Youth and Sports through the project CZ.02.2.69/0.0/0.0/16\_027/0008465.
}
\thanks{M.~Gustafsson is with the Department of Electrical and Information Technology,
Lund University, 221~00 Lund, Sweden (e-mail: mats.gustafsson@eit.lth.se).}
\thanks{M.~Capek is with the Department of Electromagnetic Field, Faculty of Electrical Engineering, Czech Technical University in Prague, 166~27 Prague, Czech Republic (e-mail: miloslav.capek@fel.cvut.cz).}
}

\maketitle

\begin{abstract}
Fundamental bounds on antenna gain are found via convex optimization of the current density in a prescribed region. Various constraints are considered, including self-resonance and only partial control of the current distribution. Derived formulas are valid for arbitrarily shaped radiators of a given conductivity. All the optimization tasks are reduced to eigenvalue problems, which are solved efficiently. The second part of the paper deals with superdirectivity and its associated minimal costs in efficiency and Q-factor. The paper is accompanied with a series of examples practically demonstrating the relevance of the theoretical framework and entirely spanning wide range of material parameters and electrical sizes used in antenna technology. Presented results are analyzed from a perspective of effectively radiating modes. In contrast to a common approach utilizing spherical modes, the radiating modes of a given body are directly evaluated and analyzed here. All crucial mathematical steps are reviewed in the appendices, including a series of important subroutines to be considered making it possible to reduce the computational burden associated with the evaluation of electrically large structures and structures of high conductivity.
\end{abstract}

\textbf{\small{\emph{Index Terms}---Antenna theory, current distribution, eigenvalues and eigenfunctions, optimization methods, directivity, antenna gain, radiation efficiency.}}

\IEEEpeerreviewmaketitle

\section{Introduction}

A question of how narrow a radiation pattern can be or, in terms of standard antenna terminology~\cite{IEEEantennaterms1993}, what are the bounds on directivity and gain, has been in the spotlight of antenna theorists’ and physicists’ for many years.

Early works studied needle-like radiation patterns~\cite{Oseen1922}. A series of works starting in the 1940s revealed the fact that the directivity is unbounded~\cite{Bouwkamp+deBruijn1946} but also predicted the enormous cost in other antenna parameters, namely in Q-factor~\cite{Harrington1960}, related sensitivity of feeding network~\cite{Yaru1951}, and radiation efficiency in case that the antenna is made of lossy material~\cite{Hansen1990}. Consequently, as pointed out by Hansen~\cite{Hansen1981}, the superdirective aperture design requires additional constraint, replacing fixed spacing in array theory~\cite{Bloch+etal1953, Uzsoky+Solymar1956}.

In order to tighten the bounds on directivity, Harrington~\cite{Harrington1958, Harrington1960} proposed a simple formula which predicts the directivity from the number of used spherical harmonics as a function of aperture size. The number of modes radiating well and the pioneering works on bounds~\cite{Harrington1965} became popular in antenna design and hold in many realistic cases, therefore, this approach demarcated the avenue of further research. Improved formula has been proposed in~\cite{Kildal+etal2017}, suggesting that, in general, the maximum directivity in the electrically small region is equal to three. The maximum directivity is studied in \cite{Margetis+etal1998} considering a given current norm. For antenna arrays, directivity bounds are shown in~\cite{Shamonina+Solymar2015}. Trade-off between maximum directivity and Q-factor for arbitrarily shaped antennas is presented in~\cite{Gustafsson+Nordebo2013}. Upper bounds for scattering of metamaterial-inspired structures are found in~\cite{Liberal+etal2014}. Recently, a composition of Huygens multipoles has been proposed~\cite{Ziolkowski2017} to increase the directivity. Notice, however, that no losses other than the radiation were assumed which re-opens the question of the actual cost of super-directivity.

Another way to limit the directional properties is a prescribed, non-zero, material resistivity of the antenna body~\cite{Karlsson2013, Arbabi+SafaviNaeini2012}. A quantity to deal with is the antenna gain, which is always bounded if at least infinitesimal losses are assumed. It may seem reasonable at this point to argue that the losses can be overcame with a concept of superconducting antennas, however, as shown in~\cite{Hansen1990}, the increase in gain with decrease of resistivity embodies slow (logarithmic) convergence. Consequently, even tiny losses, which are always present at RF, restrict the gain to a finite number.

Tightly connected is the question of maximum achievable absorption cross-section. The capability to effectively radiate energy in a certain direction can reciprocally be understood as a potential to absorb energy from that direction~\cite{Silver1949, Elliott2003}. This can be interpreted as an ability of a receiver to distort the near-field so that the incoming energy is effectively absorbed in the receiver’s body or concentrated at the receiving port. It has been realized that such an area can be huge as compared to the physical size of the particle or the physical antenna aperture~\cite{Bohren1998, Paul+Fischer1983}. Fundamental bounds on absorption cross-sections are proposed in~\cite{Sohl+etal2007c,Miller+etal2016}.

The importance to establish fundamental bounds on gain and absorption cross-section are underlined by recent development in design of superdirective (supergain) antennas and arrays~\cite{Altshuler+etal2005, Yaghjian+etal2008, Kim+etal2012, Pigeon+etal2014, Liberal+etal2015, Diao+Warnick2018}, partly fueled by the advent of novel materials and technologies~\cite{Engheta+Ziolkowski2006, Arslanagi+Ziolkowski2018}.

The procedure developed in this paper relies on convex optimization~\cite{Boyd+Vandenberghe2004} of current distributions~\cite{Gustafsson+Nordebo2013}. In order to find the optimal current distribution in a prescribed region, the antenna quantities are expressed as quadratic forms of corresponding matrix operators~\cite{Harrington1968,Gustafsson+etal2016a}. This makes it possible to solve the optimization problems rigorously via eigenvalue problems~\cite{Capek+etal2017b,TEAT-7260}. The procedure is general as arbitrarily shaped regions can be investigated. Additional constraints are enforced, \eg{}, self-resonance and restricted controllability of the current~\cite{Gustafsson+Nordebo2013,Jelinek+Capek2017}. Much work in this area has already been done in determining bounds on Q-factor~\cite{Capek+etal2017b}, radiation efficiency~\cite{Jelinek+Capek2017}, superdirectivity~\cite{Gustafsson+Nordebo2013}, gain~\cite{Harrington1965}, and capacity~\cite{Ehrenborg+Gustafsson2018}. The recent trend, followed by this paper, is to understand the mutual trade-offs between various parameters~\cite{Gustafsson2013a, Jonsson+etal2017, TEAT-7260}.

The original approach from~\cite{Harrington1965} and \cite{Harrington1968} maximizing the Rayleigh quotient for antenna gain via a generalized eigenvalue problem is recast here into an eigenvalue problem of reduced rank. Such a formulation is compatible with fast numerical methods~\cite{Chew+etal2001}, therefore, the results can be presented in a wide frequency range, \mbox{$ka \in \left[ 10^{-3}, 10^{3} \right]$}, where~$ka$ is used throughout the paper to denote the dimensionless frequency with $k$ being the wavenumber and~$a$ being the radius of a sphere circumscribing all the sources. The surface resistivity used spans the interval from extremely low values, \mbox{$\Rs=10^{-8}\,\Ohmps$}, reachable in RF superconducting cavities~\cite{Padamsee+etal1998}, through values valid for copper at RF (\mbox{$\Rs\approx 0.01\,\Ohmps$}, $f=1$\,GHz), to poor conductors of surface resistivity \mbox{$\Rs=1\,\Ohmps$}.

Optimal currents presented in this paper maximize the antenna gain. Therefore, taking reciprocity into account, they delimit the maximum effective area of any receiver designed in that region as well. For this reason, the proportionality between gain and effective area is utilized, making it possible to judge the real performance of designed and manufactured antennas, arrays, scatterers, and other radiating systems.

The behavior of the optimal solution evolves markedly with electrical size. Huygens source formed by electric and magnetic dipoles is strictly preferred in electrically small (sub-wavelength) region and a large effect of self-resonance, if enforced, is observed. End-fire radiation and negligible effect of self-resonance constraint is observed in an intermediate region. Finally, broadside radiation dominates in the electrically large region with the effective area being proportional to the cross-section area.

The paper is organized as follows. Antenna gain and effective area are introduced in Section~\ref{S:Definitions} and expressed as quadratic forms in the currents. The optimal currents are then found for maximum gain in Sections~\ref{S:maxGainT} and \ref{S:maxGsr}, including cases with additional constraints like self-resonance. Examples covering various aspects  of antenna design are presented in Section~\ref{S:maxGAeffEx}. Superdirective currents are found in Section~\ref{S:superdir} and presented as a trade-off between required directivity and minimum ohmic losses or Q-factor. All presented examples reveal the enormous cost of superdirectivity. The maximum gain is reinterpreted in Section~\ref{S:ModesAndDoF} in terms of number of sufficiently radiating modes of a structure and the results are linked back to Harrington's formula. The paper is concluded in Section~\ref{S:Conclusion}. All required mathematical tools are reviewed and key derivations are presented in the Appendices.

\section{Gain and Effective Area}
\label{S:Definitions}
Antenna gain describes how an antenna converts input power into radiation in a specified direction~$\rvh$, \cite{Balanis2012}. The gain in a direction~$\rvh$ is determined as~$4\pi$ times the quotient between the radiation intensity~$P(\rvh)$ and the dissipated power~$\Prad+\Pl$,
\begin{equation}
  G(\rvh) = 4\pi\frac{P(\rvh)}{\Prad+\Pl},
\label{eq:gain}
\end{equation}
where $\Prad$ and $\Pl$ denote the radiated power and power dissipated in ohmic and dielectric losses, respectively. The effective area, $\Aeff$, is an alternative quantity used to describe directive properties for receiving antennas, which is for reciprocal antennas simply related to the gain as~\cite{Silver1949}
\begin{equation}
  \Aeff = \frac{G\lambda^2}{4\pi},
\label{eq:Aeff}
\end{equation}
where~$\lambda=2\pi/k$ denotes the wavelength. It is seen that maximization of gain is equivalent to maximization of effective area~\cite{Harrington1968}. 

The optimized parameters are expressed in the current density~$\Jv(\rv)$ which is expanded in a set of basis functions $\left\{\basv_n\left(\rv\right)\right\}$ as~\cite{Harrington1968}
\begin{equation}
\label{eq:basFcns}
\Jv(\rv) \approx \sum_{n=1}^{N} I_n\basv_n(\rv),
\end{equation}
where the expansion coefficients, $I_n$, are collected in the column matrix $\Jm$.
This substitution yields algebraic expressions for radiation intensity, radiated power, and ohmic losses as
follows~\cite{Gustafsson+etal2016a}
\begin{align}
\label{eq:Pr}
P(\rvh) & \approx \displaystyle\frac{1}{2} \left| \Fm\Jm \right|^2 = \frac{1}{2}\Jm^\herm\Fm^\herm\Fm\Jm, \\
\label{eq:Prad}
\Prad & \approx \displaystyle\frac{1}{2} \Jm^{\herm}\Rmr\Jm, \\
\label{eq:Ploss}
\Pl & \approx \displaystyle\frac{1}{2} \Jm^{\herm}\Rml\Jm.
\end{align}
The matrices used in \eqref{eq:Pr}--\eqref{eq:Ploss} are reviewed in Appendix~\ref{S:MoMmatrices}. Substitution of \eqref{eq:Pr}--\eqref{eq:Ploss} into~\eqref{eq:gain} yields
\begin{equation}
  G(\rvh) \approx 4\pi\frac{|\Fm\Jm|^2}{\Jm^{\herm}(\Rmr+\Rml)\Jm}
  =  4\pi\frac{\Jm^{\herm}\Um\Jm}{\Jm^{\herm}(\Rmr+\Rml)\Jm},
\label{eq:gainMat}
\end{equation}
where we also introduced the matrix $\Um=\Fm^{\herm}\Fm$ to simplify the notation and highlight the expression of the gain~$G(\rvh)$ as a Rayleigh quotient.

\subsection{Maximum Gain: Tuned Case}
\label{S:maxGainT}

The maximum gain for antennas confined to a region~\mbox{$\rv \in \reg$} is formulated as the optimization problem
\begin{equation}
\begin{aligned}
	& \maximize && \Jm^{\herm}\Um\Jm \\
	& \subto &&  \Jm^{\herm}(\Rml+\Rmr)\Jm = 1,
\end{aligned}
\label{eq:gain_res}
\end{equation}
where for simplicity the dissipated power is normalized to unity. This problem is equivalent to the Rayleigh quotient
\begin{equation}
  \Gmax \approx 4\pi \max_{\Jm} \frac{\Jm^{\herm}\Fm^{\herm}\Fm\Jm}{\Jm^{\herm}(\Rmr+\Rml)\Jm},
\label{eq:maxgain1}
\end{equation}
to which a solution is found via the generalized eigenvalue problem~\cite{Harrington1968}
\begin{equation}
  \Fm^{\herm}\Fm\Jm = \eigv(\Rmr+\Rml)\Jm.
\label{eq:Geigorig}
\end{equation}

In order to reduce the computational burden, the formula~\eqref{eq:Geigorig} is further transformed to
\begin{equation}
  \left( \Rmr+\Rml \right)^{-1}\Fm^{\herm}\Fm\Jm = \eigv\Jm
\end{equation}
and multiplied from left by the matrix~$\Fm$. By introducing~\mbox{$\Jmt=\Fm\Jm$} we readily get
\begin{equation}
  \Fm(\Rmr+\Rml)^{-1}\Fm^{\herm}\Jmt = \eigv\Jmt.
  \label{eq:maxgain_dual}
\end{equation}
Taking into account that the far-field matrix~$\Fm$ can be expressed using two orthogonal polarizations, see Appendix~\ref{S:MoMmatrices}, the original~$N\times N$ eigenvalue problem~\eqref{eq:Geigorig} is reduced into the~$2\times 2$ eigenvalue problem~\eqref{eq:maxgain_dual} which can be written as
\begin{equation}
  \Gmax \approx 4\pi \max\eig(\Fm(\Rmr+\Rml)^{-1}\Fm^{\herm})
\label{eq:maxgain}
\end{equation}
with the optimal current determined as
\begin{equation}
  \Jm = \eigv^{-1}(\Rmr+\Rml)^{-1}\Fm^{\herm}\Jmt.
\label{eq:maxgainI}
\end{equation}
The corresponding case with the partial gain contains one polarization direction and hence the eigenvalue \mbox{$\eigv=\Fm(\Rmr+\Rml)^{-1}\Fm^{\herm}$} and current
\begin{equation}
  \Jm \sim (\Rmr+\Rml)^{-1}\Fm^{\herm}.
\label{eq:maxgainPhaseCorr}
\end{equation}
Here, the~$\Fm^{\herm}$ part can be interpreted as phase conjugation of an incident plane wave from the~$\rvh$-direction, and hence the current corresponding to the maximum gain is found by phase conjugation of the incident wave modified by~$(\Rmr+\Rml)^{-1}$.    

\subsection{Maximum Gain: Self-Resonant Case}
\label{S:maxGsr}

The solution to~\eqref{eq:maxgain} is in general not self-resonant. Self resonance is enforced to~\eqref{eq:maxgain} by adding the constraint of zero reactance, $\Jm^{\herm}\Xm\Jm=0$, see Appendix~\ref{S:MoMmatrices}, producing the optimization problem 
\begin{equation}
\begin{aligned}
	& \maximize && \Jm^{\herm}\Um\Jm \\
	& \subto &&  \Jm^{\herm}\Xm\Jm = 0 \\
	& && \Jm^{\herm}(\Rml+\Rmr)\Jm = 1.
\end{aligned}
\label{eq:maxgain_res}
\end{equation}
This optimization problem is a \ac{QCQP}, see Appendix~\ref{S:QCQP}, that is transformed to a dual problem by multiplication of $\Jm^{\herm}\Xm\Jm$ with a scalar parameter~$\nu$ and adding the constraints together, \ie{},
\begin{equation}
\begin{aligned}
	& \maximize && \Jm^{\herm}\Um\Jm \\
	& \subto && \Jm^{\herm}(\nu\Xm+\Rml+\Rmr)\Jm = 1,
\end{aligned}
\label{eq:maxgain_res1}
\end{equation}
which is solved as a generalized eigenvalue problem analogously to Section~\ref{S:maxGainT}. The solution to this problem is greater or equal to~\eqref{eq:maxgain_res} and taking its minimum value produces the dual problem~\cite{Boyd+Vandenberghe2004}
\begin{equation}
\begin{aligned}
  \Gmaxr & \approx 4\pi \min_{\nu} \max \eig(\Um,\nu\Xm+\Rml+\Rmr) \\
         & = 4\pi \min_{\nu} \max \eig(\Fm(\nu\Xm+\Rml+\Rmr)^{-1}\Fm^{\herm})
\label{eq:maxgain_res_dual}
\end{aligned} 
\end{equation}
which is convex and easy to solve, \eg, with the bisection algorithm~\cite{Nocedal+Wright2006}. The derivative of the eigenvalue~$\gamma$ with respect to~$\nu$ is~\cite{Lancaster1964}
\begin{equation}
  \partder{\eigv}{\nu}
  =-\eigv^2\frac{\Jm^{\herm}\Xm\Jm}{\Jm^{\herm}\Um\Jm} 
   \begin{cases}
  \leq 0 \qtext{for } \nu\leq \nuopt, \qtext{inductive} \\
  = 0 \qtext{for } \nu =  \nuopt, \qtext{resonant} \\
  \geq 0 \qtext{for } \nu\geq \nuopt, \qtext{capacitive}
  \end{cases}
\label{eq:eigderivative}
\end{equation}
for cases with non-degenerate eigenvalues. Degenerate eigenvalues are often related to geometrical symmetries and solved by decomposition of the current~$\Jm$ into orthogonal sub-spaces~\cite{Capek+etal2017b}. The range~\mbox{$\nu \in \left[\nu_\mrm{min}, \nu_\mrm{max}\right]$} in~\eqref{eq:maxgain_res_dual} is determined from the condition $\nu\Xm+\Rml+\Rmr\succeq\Om$ which can be computed from the smallest and largest eigenvalues, $\eig(\Xm,\Rml+\Rmr)$, \ie,
\begin{equation}
  \frac{-1}{\max\eig(\Xm,\Rml+\Rmr)}
  \leq \nu \leq 
  \frac{-1}{\min\eig(\Xm,\Rml+\Rmr)},
\label{eq:nurange}
\end{equation} 
see Appendix~\ref{S:nuregion} for details.

The minimal eigenvalue~$\min\eig(\Xm,\Rml+\Rmr)$ is related to the Q-factor of the maximal capacitance in the geometry which is very large for all considered cases giving an upper limit very close to zero and \mbox{$\nu_\mathrm{max} \rightarrow 0$} as the mesh is refined. The maximal eigenvalue is related to the maximal inductive Q-factor which is a fixed value depending on shape of the object and gives the lower bound $\nu_\mathrm{min}$, \cf{} with the inductor Q-factor in~\cite{TEAT-7260}. 

\subsection{Examples of Maximum Gain and Effective Area}
\label{S:maxGAeffEx}

The following section presents maximum gain and effective area for examples of various complexity:
\begin{enumerate}
    \item spherical shell both for externally and self-resonant currents, Section~\ref{S:maxGAeffEx1},
    \item comparison of end-fire and broadside radiation from a rectangular region, Section~\ref{S:maxGAeffEx2},
    \item maximization of effective area if different parts of a cylinder are considered, Section~\ref{S:maxGAeffEx3},
    \item limited controllability of currents for a parabolic dish with spherical prime feed, Section~\ref{S:maxGAeffEx4}.
\end{enumerate}

\subsubsection{Externally tuned and self-resonant currents (spherical shell)}
\label{S:maxGAeffEx1}

Expansion of the current density on a spherical shell in vector spherical harmonics~\cite{Kristensson2016} produces diagonal reactance~$\Xm$, radiation resistance~$\Rmr$, and loss~$\Rml$ matrices with closed form expressions of the elements. The direction of radiation can without loss of generality be chosen to~$\rvh=\zvh$ for which the elements~$\Fm$ are zero for azimuthal Fourier indices~$|m|\neq 1$. It is hence sufficient to consider~$|m|=1$ for the radiation, see Appendix~\ref{S:sphere}. 

\begin{figure}
\includegraphics[width=\columnwidth]{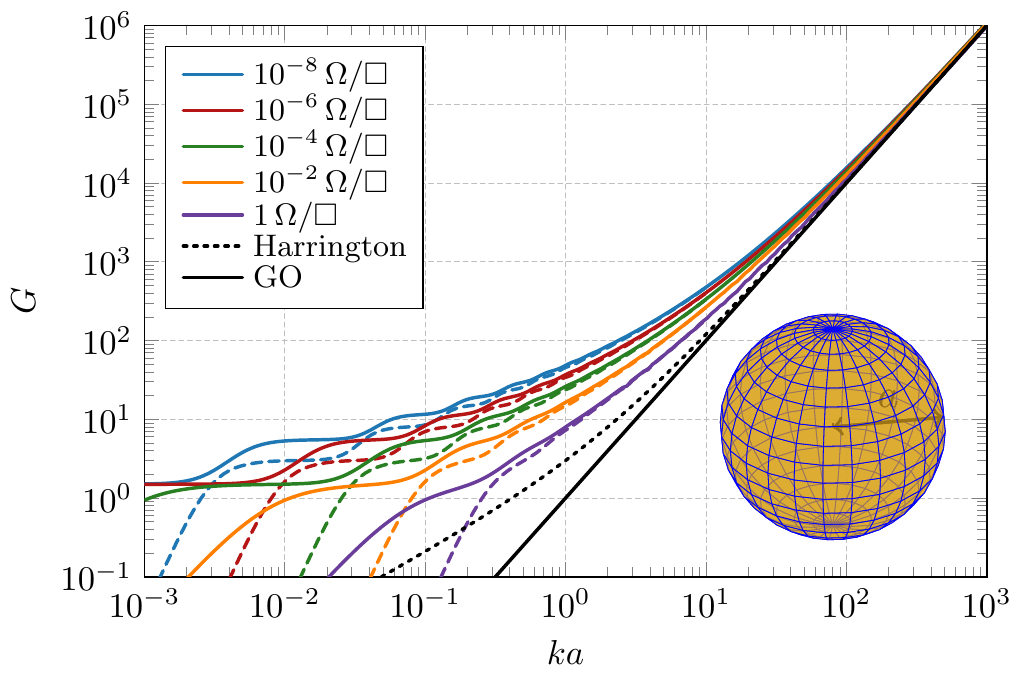}
\caption{Maximum gain for a spherical shell of radius~$a$ with surface resistivity~$\Rs=10^{-n}\Ohmps$, \mbox{$n=\left\{0,2,4,6,8\right\}$}, both for externally tuned~\eqref{eq:maxgain}, $\Gmax$, (solid lines) and for self-resonant~\eqref{eq:maxgain_res_dual}, $\Gmaxr$, (dashed lines) currents.}
\label{fig:SphereGain}
\end{figure}

\begin{figure}
\centering
\includegraphics[width=88mm]{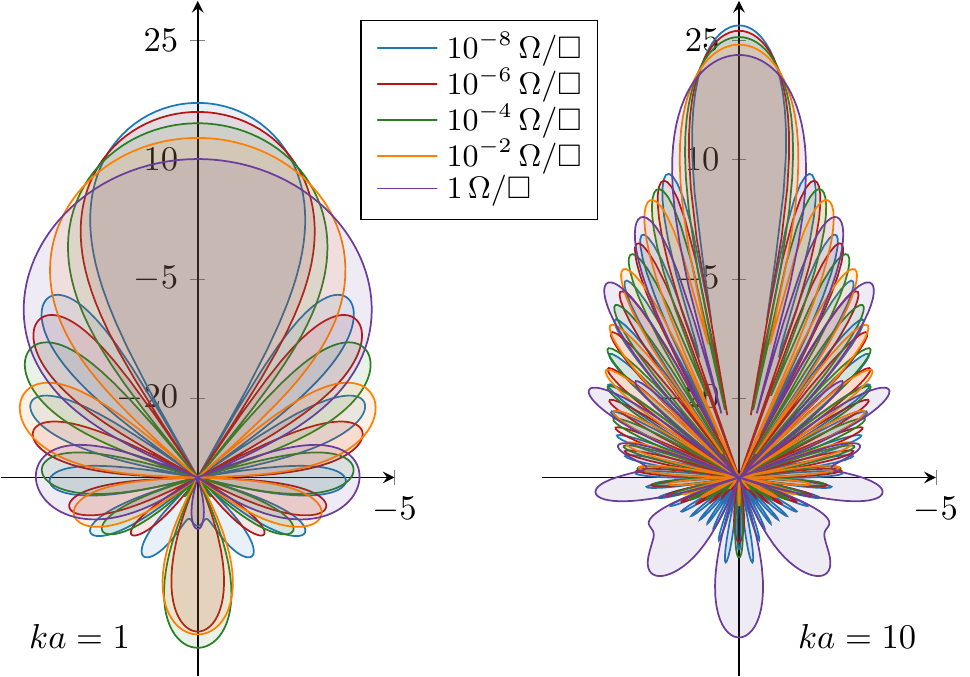}
\caption{Radiation patterns for a spherical shell of radius~$a$ with surface resistivity~$\Rs=10^{-n}\Ohmps$, \mbox{$n=\left\{0,2,4,6,8\right\}$} corresponding to the externally tuned case in Fig.~\ref{fig:SphereGain}. The radiation patterns are shown in terms of gain $G$ for a~$\vartheta$-cut and~$\varphi=0$. The two electrical sizes, \mbox{$ka =1$} and \mbox{$ka = 10$}, are depicted.}
\label{fig:SphRadPat}
\end{figure}

\begin{figure}
\includegraphics[width=\columnwidth]{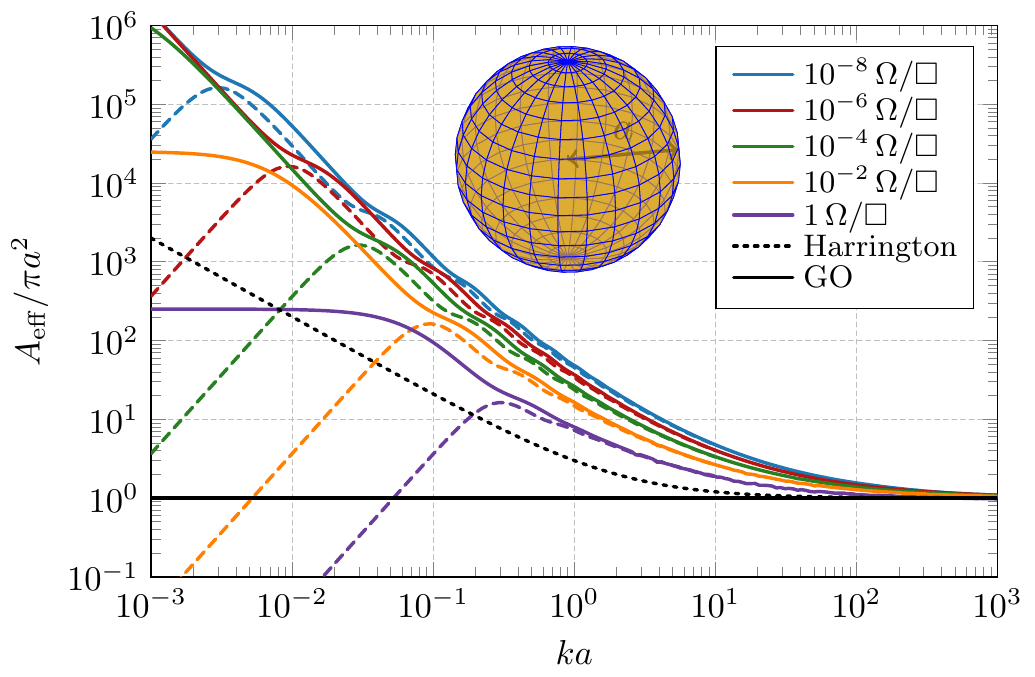}
\caption{Maximum effective area for a spherical shell of radius~$a$ with surface resistivity~$\Rs=10^{-n}\Ohmps$, \mbox{$n=\left\{0,2,4,6,8\right\}$}, both for externally tuned~\eqref{eq:maxgain} (solid lines) and for self-resonant~\eqref{eq:maxgain_res_dual} (dashed lines) currents.}
\label{fig:SphereAeff}
\end{figure}

The maximum gain for a spherical shell with surface resistivity~\mbox{$\Rs=10^{-n}\Ohmps$} for \mbox{$n=\left\{0,2,4,6,8\right\}$} is determined using~\eqref{eq:maxgain}, \eqref{eq:maxgain_res_dual} and depicted in Fig.~\ref{fig:SphereGain}. The results are compared with the estimates~\mbox{$G_{\mrm{H}}=(ka)^2+2ka$} by Harrington~\cite{Harrington1960} and from the geometrical cross section~\mbox{$G_{\mrm{GO}}=4\pi\Across/\lambda^2$}. It is observed that the additional constraint on self-resonance, \ie{},~\mbox{$\Jm^{\herm}\Xm\Jm = 0$}, in~\eqref{eq:maxgain_res} has a large effect for small structures ($ka<1$) but negligible effect for electrically large structures. The tuned and self-resonant cases have $D=3/2$ and $D=3$, respectively, in the limit of electrically small structures ($ka\to 0$), see Appendix~\ref{S:sphere}. Onset of spherical modes for small~$ka$ gives a step-wise increasing directivity and gain, see figures in Appendix~\ref{S:sphere}. Dependence on~$\Rs$ diminishes and the gain approaches $G_{\mrm{GO}}$ as $ka$ increases. The radiation patterns and the influence of the surface resistivity on the maximum gain~$G$ is shown in Fig.~\ref{fig:SphRadPat} for the externally tuned case and electrical sizes~$ka=1$ and $ka=10$.
The electrically large limit is more clearly seen by plotting the effective area~\eqref{eq:Aeff} in Fig.~\ref{fig:SphereAeff}, where it is observed that the effective area approaches the cross-section area as $ka\to\infty$.

\subsubsection{Broadside and end-fire radiation (rectangular plate)}
\label{S:maxGAeffEx2}

\begin{figure}%
\includegraphics[width=\columnwidth]{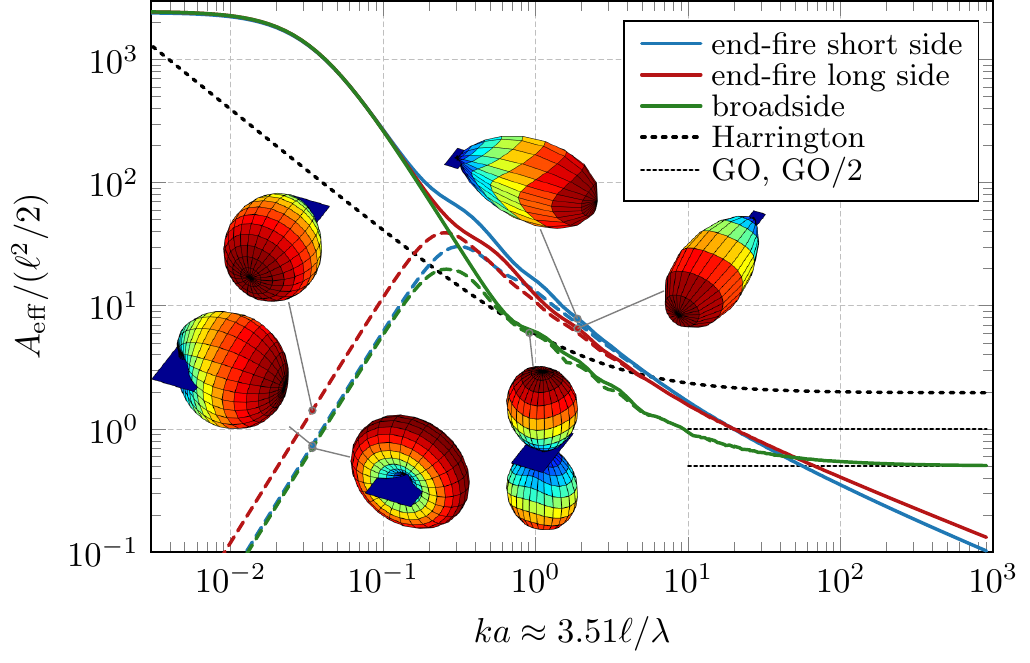}
\caption{Maximum effective area in the cardinal directions for a rectangular plate with size~$\ell\times\ell/2$ and surface resistivity~\mbox{$\Rs=10^{-4}\ZVAC$} per square. Bounds for externally tuned~\eqref{eq:maxgain} (solid lines) and self-resonant~\eqref{eq:maxgain_res} (dashed lines) currents are depicted.}
\label{fig:RecAeff}
\end{figure}

The symmetry of the sphere is ideal for analytic solution of the optimization problem but cannot be used to investigate important cases such as broadside and end-fire radiation~\cite{Elliott2003}. Let us, therefore, consider a planar rectangular plate with side lengths~$\ell$ and~\mbox{$\ell/2$} placed at~$z=0$ having surface resistivity~\mbox{$\Rs=10^{-4}\ZVAC$} per square. The maximum effective area is depicted in Fig.~\ref{fig:RecAeff} for radiation in the cardinal directions. Three regions can be identified: electrically small ($ka\ll 1$) with large difference between the externally tuned and self-resonant cases, intermediate region with dominant end-fire radiation, and electrically large $ka\gg 1$ with dominant broadside radiation. 

Negligible directional differences are observed for the electrically small ($ka\ll 1$) externally tuned case which can be explained by radiation patterns originating from electric dipoles. The effective area for the self-resonant case deceases as $(ka)^2$ and consists of a combination of electric and magnetic dipoles. Huygens sources are obtained for the end-fire cases where the gain is higher for radiation along the longest side than for the shorter side due to its lower amount of stored electric energy. Gain in the broadside direction is lower due to its up-down symmetric radiation pattern. 

The difference between the externally tuned and self-resonant cases decreases as $ka$ increases and become negligible around $ka\approx 1$. Here, it is also seen that the effective area for the self-resonant case has a maximum around the same size. The end-fire directions have higher effective area (and gain) than the broadside direction up to \mbox{$ka\approx 50$}. Approaching the electrically large region ($ka\to\infty$), the broadside radiation converges to one half of the physical cross section area since the electric currents produce symmetric radiation patterns in up-down direction, and the end-fire directions are observed to decay approximately as~$(ka)^{-1/2}$.

\subsubsection{Contribution to the maximum effective area (cylinder)}
\label{S:maxGAeffEx3}

The maximum effective area is studied in this example for a single disc~$\reg_{\mrm{t}}$, two separated discs~$\reg_{\mrm{t}}\cup\reg_{\mrm{b}}$, a mantel surface~$\reg_{\mrm{m}}$ and a cylinder~$\reg_{\mrm{t}}\cup\reg_{\mrm{b}}\cup\reg_{\mrm{m}}$.

\begin{figure}
\includegraphics[width=\columnwidth]{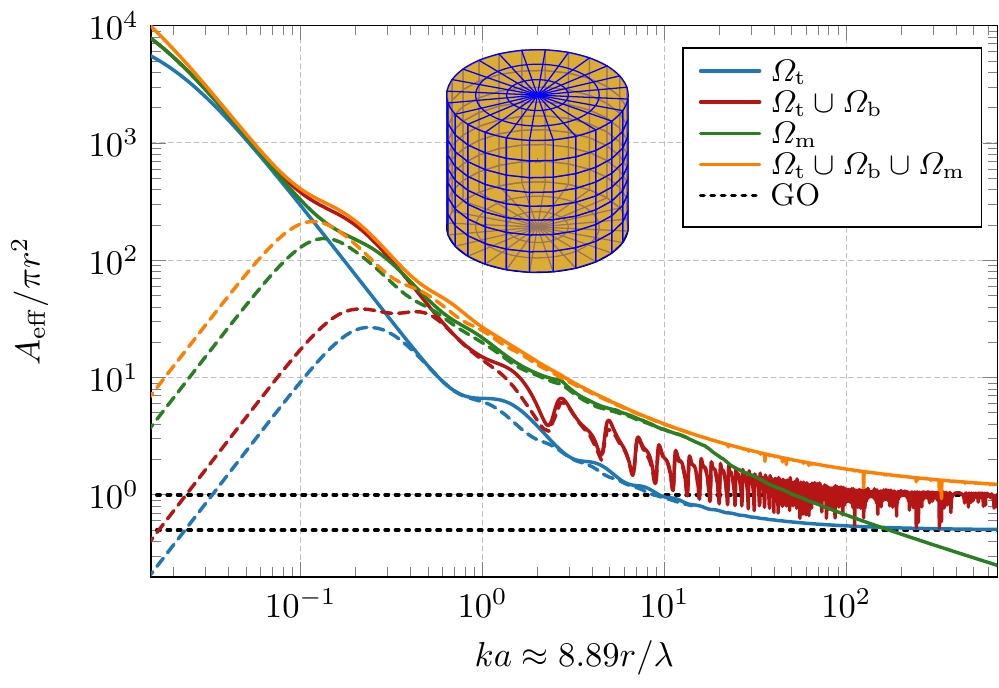}
\caption{Maximum effective area for a disc, two discs, a mantel surface, and a cylindrical structure with surface resistivity $\Rs=10^{-2}\Ohmps$ in the axial ($\zvh$)-direction. Bounds for externally tuned~\eqref{eq:maxgain} (solid lines) and self-resonant~\eqref{eq:maxgain_res} (dashed lines) currents are depicted.}
\label{fig:CylAeff}
\end{figure}

The performance of a single disc~$\reg_{\mrm{t}}$ with radius~$r$ depicted in Fig.~\ref{fig:CylAeff} confirms the broadside limit~\mbox{$\Aeff\to \Across/2$} in the electrically large region as observed for the rectangle in Fig.~\ref{fig:RecAeff}. The stepwise decrease for smaller sizes can be interpreted as the onset of spherical modes in agreement with the sphere in Fig.~\ref{fig:SphereGain}. Addition of a second disc separated by the distance~$2 r$ from the first disc breaks the up-down symmetry of the radiation pattern. The effective area is depicted~in Fig.~\ref{fig:CylAeff} with the curve labeled~$\reg_{\mrm{t}}\cup\reg_{\mrm{b}}$. A rapid oscillatory behavior is observed for electrically large structures. These oscillations are due to the up-down symmetry for disc distances of integer multiples of the wavelength, \ie, the radiation in the~$\pm\zvh$-directions are identical, where~$\zvh$ denotes the axis of rotation. For other distances the radiation from the discs can contribute constructively in the $\zvh$-direction and destructively in the $-\zvh$-direction. This produces an effective area approaching~$\Across$ on average in the electrically large ($ka\to\infty$) region. 

End-fire radiation is considered from the mantel surface~$\reg_{\mrm{m}}$ (the hollow cylindrical structure without top and bottom discs) in Fig.~\ref{fig:CylAeff}. The effective area decreases approximately linearly in the log-log scale giving the approximate scaling~\mbox{$\Aeff\sim (ka)^{-1/2}$} as also seen in Fig.~\ref{fig:RecAeff}. Here, it is also observed that the effect of resistivity is larger for the end-fire case as compared to the broadside cases. 

Adding the bottom and top discs to the cylinder mantel surface  forms a cylindrical shell as shown in Fig.~\ref{fig:CylAeff}. The effective area approaches~$\Across$ similar to the discs case but with most of the oscillations removed. 

\subsubsection{Controllable currents (parabolic reflector)}
\label{S:maxGAeffEx4}

\begin{figure}
\includegraphics[width=\columnwidth]{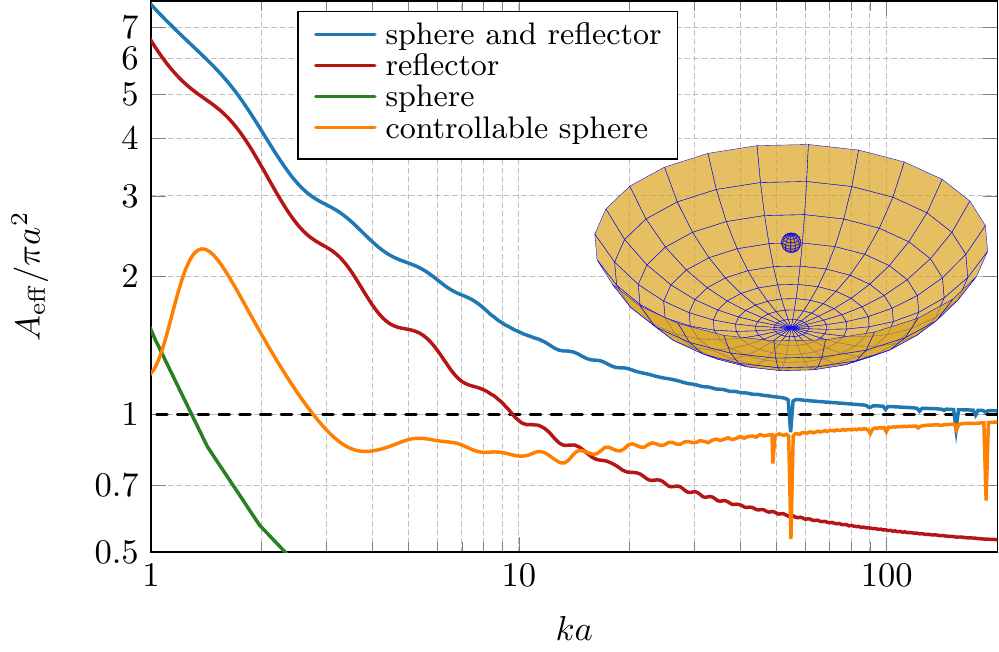}
\caption{Maximum effective area from~\eqref{eq:maxgain} for a parabolic reflector combined with a sphere placed in the focal point with surface resistivity $\Rs=10^{-2}\Ohmps$ in the axial ($\zvh$)-direction. The parabolic reflector has radius~$a$, focal distance~$a/2$, and depth~$a/2$ and the sphere has radius \mbox{$r=a/20$}.}
\label{fig:ParaAeff}
\end{figure}

A parabolic reflector is used to illustrate the effective area for controllable substructures, see Fig.~\ref{fig:ParaAeff}. The parabolic reflector is rotationally symmetric and has radius~$a$, focal distance~$a/2$, and depth~$a/2$. A sphere with radius \mbox{$r=a/20$} is placed in the focal point. Maximum effective area is depicted for three cases: controllable currents on the parabolic reflector and sphere, controllable currents on the reflector, and controllable currents on the sphere. The induced currents are determined from the \ac{MoM} impedance matrix~\cite{Gustafsson+Nordebo2013}. Controlling both the reflector and sphere gives the largest effective area and approaches the cross section area for electrically large structures as seen in Fig.~\ref{fig:ParaAeff}. The oscillations starting around~$ka\approx 55$ originates in the internal resonances of the sphere, where it is noted that~$kr \approx 2.74$ in agreement with the TE dipole resonance~\cite{Harrington1961}. This is also confirmed by negligible impact on the overall behavior of the effective area from using smaller and larger spheres except for shifting of the resonances up and down. However, the scenario with both reflector and the prime feeder controllable is unrealistic.

Removing the sphere and optimizing the currents on the reflector lowers the effective area with approximately a factor of two for large~$ka$. This might at first seem surprising as the cross-section area of the reflector is~$400$ times larger than for the sphere having radius~$r=a/20$. Moreover, the effective area of the sphere is close to its cross-section area, \ie, \mbox{$\Aff\approx\pi r^2 \approx \Across/400$} as seen in Fig.~\ref{fig:SphereAeff}. The effective area of the reflector is better explained by its similarity to the disc in Fig.~\ref{fig:CylAeff} and rectangle in Fig.~\ref{fig:RecAeff}, where the asymptotic limit $\Across/2$ is explained by the up-down symmetry of the radiation pattern. The limit~$\Aff\approx\Across$ for the reflector together with the sphere is hence explained by elimination of the backward radiation.  

Replacing the controllable currents on the reflector with induced currents from the sphere produces an effective area just below~$\Across$ for high~$ka$. The reduction for small~$ka$ is similar to the short circuit of the currents above a ground plane. Internal resonances for the sphere are more emphasized as all radiation originates from the sphere in this case.

\section{Superdirectivity}
\label{S:superdir}

Directional properties of the radiation pattern are quantified by the directivity
\begin{equation}
  D(\rvh) = 4\pi\frac{P(\rvh)}{\Prad}
  \approx 4\pi\frac{\Jm^{\herm}\Um\Jm}{\Jm^{\herm}\Rmr\Jm}.
\label{eq:D}
\end{equation}
Here, it is seen that the directivity~\eqref{eq:D} only differs from the gain~\eqref{eq:gain} by its normalization with the radiated power instead of the total dissipated power. This difference is the radiation efficiency $\eta=\Prad/(\Prad+\Pl)$, which is related to the dissipation factor
\begin{equation}
  \delta = \frac{\Pl}{\Prad}
  \approx \frac{\Jm^{\herm}\Rml\Jm}{\Jm^{\herm}\Rmr\Jm}.
\label{eq:delta}
\end{equation}
Directivity higher than a nominal directivity is often referred to as superdirectivity and associated with low efficiency and narrow bandwidth~\cite{Hansen1981}. The trade-off between the Q-factor and directivity was shown in~\cite{Gustafsson+Nordebo2013} and further investigated in~\cite{Gustafsson+etal2016a,Jonsson+etal2017,Gustafsson+etal2015b}. Superdirectivity is also associated with decreased radiation efficiency or equivalently an increased dissipation factor~\eqref{eq:delta}.

\subsection{Trade-off Between Dissipation Factor and Directivity}

The trade-off between losses and directivity for a self-resonant antenna can be analyzed by separating the radiated power~$\Prad$ and losses~$\Pl$ in~\eqref{eq:maxgain_res} giving the optimization problem
\begin{equation}
\begin{aligned}
	& \maximize && \Jm^{\herm}\Um\Jm\\
	& \subto &&  \Jm^{\herm}\Xm\Jm = 0 \\  
	& && \Jm^{\herm}\Rmr\Jm = 1\\
	& && \Jm^{\herm}\Rml\Jm = \delta.
\end{aligned}
\label{eq:maxGD_opt}
\end{equation}
The constraint $\Jm^{\herm}\Xm\Jm=0$ is dropped for the corresponding non-self resonant case~\eqref{eq:gainMat}. The Pareto front is formed by adding the constraints weighted by scalar parameters, \ie{},
\begin{equation}
\begin{aligned}
	& \maximize && \Jm^{\herm}\Um\Jm\\
	& \subto &&  \Jm^{\herm}(\nu\Xm+\alpha\Rml+\Rmr)\Jm = 1,
\end{aligned}
\label{eq:maxGD_pareto}
\end{equation}
where the right-hand side is re-normalized to unity without restriction of generality. This problem is identical to the maximum gain problem~\eqref{eq:maxgain_res1} if the Pareto parameter~$\alpha\geq 0$ is included in the surface resistivity~$\Rs$ and is hence solved as the eigenvalue problem~\eqref{eq:maxgain_res_dual}. 
Here,~$\alpha=0$ solely weights the radiated power regardless of ohmic losses and increasing~$\alpha$ starts to emphasize ohmic losses. The maximal directivity ($\alpha=0$) is in general unbounded~\cite{Oseen1922,Bouwkamp+deBruijn1946} but has low gain. The other extreme point $\alpha\to\infty$ neglects the radiated power and maximizes $D/\delta$, \ie, the quotient between the directivity and dissipation factor.   
\begin{figure}
\includegraphics[width=\columnwidth]{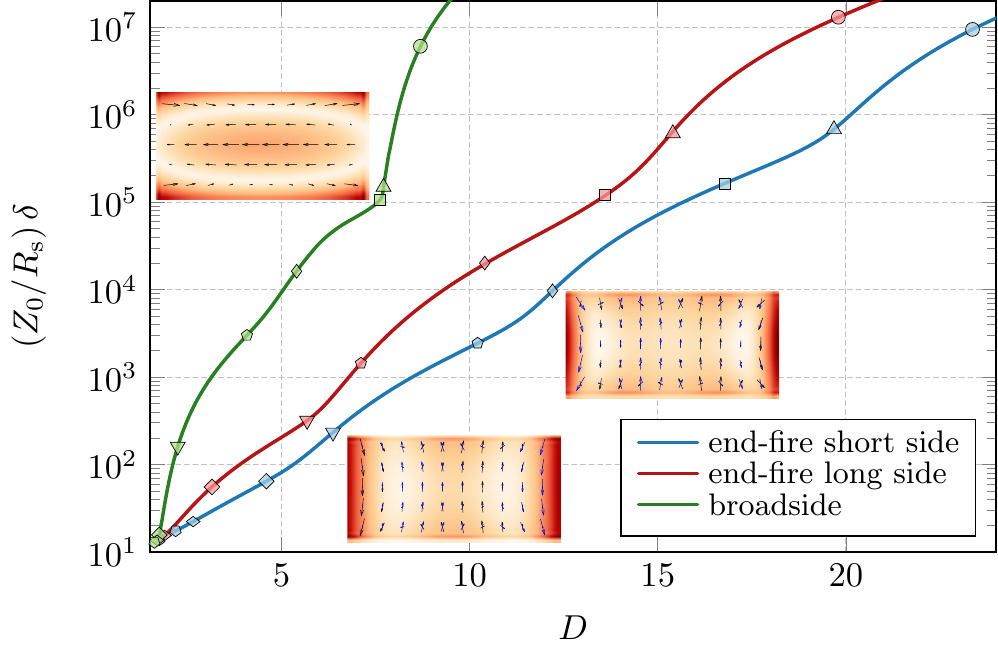}
\caption{Minimum externally tuned dissipation factor for a rectangular plate of side aspect ratio~$2:1$ and electrical size~$ka=1$ as a function of directivity~$D$ in the cardinal directions. The corresponding case with maximum gain is depicted in Fig.~\ref{fig:RectangleGvsR}.}
\label{fig:RectangleDeltaVsD}
\end{figure}

\begin{figure}
\includegraphics[width=\columnwidth]{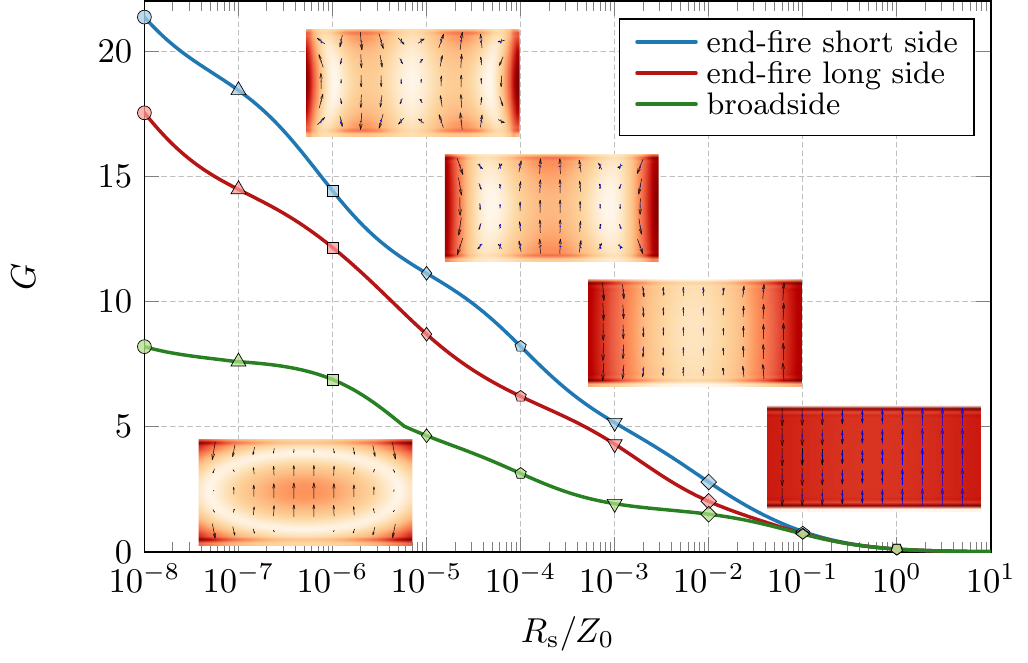}%
\caption{Maximum externally tuned gain for a rectangular plate of side aspect ratio~$2:1$ and electrical size~$ka=1$ as a function of surface resistivity~$\Rs$. The current density is depicted for~$\Rs\in\{10^{-6},10^{-4},10^{-2},1\}\ZVAC$ and $\Rs=10^{-5}\ZVAC$ for radiation in end-fire short side and broadside directions, respectively, see also Fig.~\ref{fig:RectangleDeltaVsD}.}
\label{fig:RectangleGvsR}
\end{figure}

The minimum dissipation factor for the rectangular plate from Fig.~\ref{fig:RecAeff} as a function of the directivity in the cardinal directions and its corresponding case with maximum gain as a function of surface resistivity~$\Rs$ are shown in Fig.~\ref{fig:RectangleDeltaVsD} and Fig.~\ref{fig:RectangleGvsR}, respectively. 
Although the physical interpretation of these two problems is different, they are both solved using the same eigenvalue problem and have identical current densities, \ie{}, the optimal currents were found using~\eqref{eq:maxgain_dual} which is identical to~\eqref{eq:maxGD_pareto} without the~$\Xm$-term. 
Consider, \eg, the blue curve depicting end-fire radiation along the short side. The normalized dissipation factor is monotonically increasing with~$D$ from approximately~$10$ for~$D\approx 2$ to $10^7$ for~$D\approx 25$ showing that an increased directivity comes with a high cost in losses. The corresponding blue curve in Fig.~\ref{fig:RectangleGvsR} decreases monotonically with the surface resistivity~$\Rs$ from~\mbox{$G\approx 22$} for \mbox{$\Rs=10^{-8}\ZVAC$} to \mbox{$G \approx 0.1$} for \mbox{$\Rs=\ZVAC$}. The current density is depicted for~\mbox{$\Rs\in\{10^{-6},10^{-4},10^{-2}\}\ZVAC$} in Fig.~\ref{fig:RectangleGvsR} and $\Rs\in\{10^{-5},10^{-3}\}\ZVAC$ in Fig.~\ref{fig:RectangleDeltaVsD}, where it is seen that the oscillations in the current density increase for high~$D$ and low~$\Rs$. Moreover, the markers on each curve in Fig.~\ref{fig:RectangleDeltaVsD} and Fig.~\ref{fig:RectangleGvsR} correspond to points with identical current densities. Here, it is seen that the uniform spacing in Fig.~\ref{fig:RectangleGvsR} is not preserved in Fig.~\ref{fig:RectangleDeltaVsD}, \eg{}, the green curve depicting broadside radiation has two almost overlapping points around $D\approx 8$ and $(\ZVAC/\Rs)\delta\approx 10^5$. These two points also have close to orthogonal current densities as seen by the insets and correspond to cases where the eigenvalue problem~\eqref{eq:maxGD_pareto} has degenerate eigenvalues. For these cases we use linear combinations between the eigenvectors to span the Pareto curve~\cite{Capek+etal2017b}.

The minimum dissipation factor~\cite{Jelinek+Capek2017,Thal2018,TEAT-7260} is lower than the dissipation factor obtained from the $\alpha\to\infty$ case for electrically large structures. These limit cases are connected by reformulating the problem~\eqref{eq:maxGD_opt} by either minimizing the ohmic losses or maximizing the radiated power. Minimization of ohmic losses subject to fixed radiation intensity and radiated power is
\begin{equation}
\begin{aligned}
	& \minimize && \Jm^{\herm}\Rml\Jm\\
	& \subto &&  \Jm^{\herm}\Xm\Jm = 0 \\  
	& && \Jm^{\herm}\Um\Jm = 2P\\
	& && \Jm^{\herm}\Rmr\Jm = 2 \Prad,
\end{aligned}
\label{eq:minPl_opt}
\end{equation}
which is relaxed to
\begin{equation}
\begin{aligned}
	& \minimize && \Jm^{\herm}\Rml\Jm\\
	& \subto &&  \Jm^{\herm}(\nu\Xm+\alpha\Um+\Rmr)\Jm = 1,
\end{aligned}
\label{eq:minPl_pareto}
\end{equation}
where again the right-hand side is re-normalized to unity. 

\subsection{Trade-off Between Q-factor and Directivity}

Superdirectivity is also associated with narrow bandwidth and high Q-factor~\cite{Gustafsson+Nordebo2013,Jonsson+etal2017}. Adding constraints on the stored energy to the optimization problem~\eqref{eq:maxGD_opt} results in the optimization problem 
\begin{equation}
\begin{aligned}
	& \maximize && \Jm^{\herm}\Um\Jm\\
	& \subto &&  \Jm^{\herm}\Xm\Jm =  0\\  
	& && \Jm^{\herm}(\Xme+\Xmm)\Jm =  2Q\\
	& && \Jm^{\herm}\Rml\Jm = \delta\\
	& && \Jm^{\herm}\Rmr\Jm = 1,
\end{aligned}
\label{eq:maxGQD}
\end{equation}
where $\Xme+\Xmm=k\partial\Xm/\partial k$ are matrices used to determine the stored energy~\cite{Vandenbosch2010,Gustafsson+etal2016a}.  
Forming linear combinations between the constraints is used to determine the Pareto front and analyzing the trade-off between directivity, Q-factor, and dissipation factor. 

Although~\eqref{eq:maxGQD} can be used to analyze the trade-off, it is illustrative to focus on the constraints on the dissipation factor and Q-factor separately. Dropping the constraint on the ohmic losses reduces~\eqref{eq:maxGQD} to the problem of lower bounds on the Q-factor for a given directivity~\cite{Gustafsson+Nordebo2013} which is relaxed to
\begin{equation}
\begin{aligned}
	& \maximize && \Jm^{\herm}\Um\Jm\\
	& \subto && \Jm^{\herm}(\nu\Xm+\alpha(\Xme+\Xmm)+\Rmr)\Jm = 1,
\end{aligned}
\label{eq:maxQdf_pareto}
\end{equation}
and solved analogously to~\eqref{eq:maxgain_res_dual} for fixed $\alpha$. Here, $\alpha=0$ solely weights the radiated power regardless of ohmic losses and increasing $\alpha$ starts to emphasize ohmic losses. The maximal directivity ($\alpha=0$) is in general unbounded~\cite{Oseen1922,Bouwkamp+deBruijn1946} but has a high Q-factor. Here, reformulations similar to~\eqref{eq:minPl_opt} can be used to reach the lower bound on the Q-factor. 

\begin{figure}%
\includegraphics[width=\columnwidth]{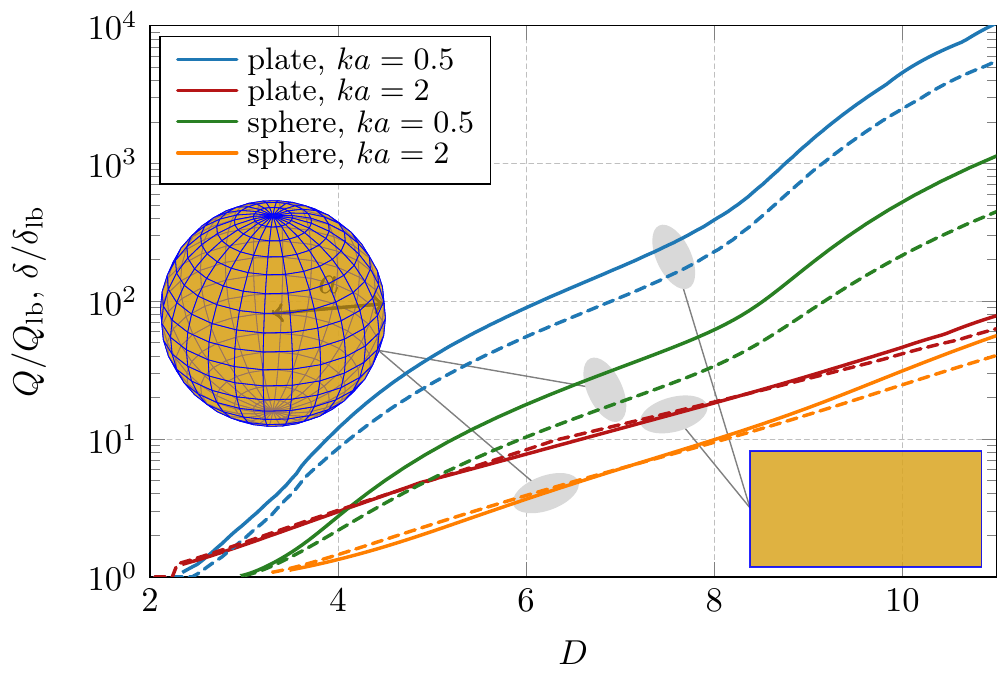}
\caption{Lower bounds on dissipation (solid lines) and Q-factors (dashed lines) for prescribed directivity~$D$ normalized with respect the lower bounds. The results were calculated for a spherical shell of radius~$a$ and a rectangular plate of side aspect ratio~$2:1$. The electrical size used is~\mbox{$ka\in\{0.5,2\}$} and the currents are self-resonant.}
\label{fig:Qdf_comp}
\end{figure}

The trade-offs between directivity and dissipation factor and Q-factor are compared in Fig.~\ref{fig:Qdf_comp} for a spherical shells and a rectangular plate of size~\mbox{$ka\in\{0.5,2\}$}. The bounds are normalized with the lower bounds on the dissipation and Q-factors for the structure. The stored energy matrices are transformed to be positive semidefinite for the Q-factor calculation~\cite{Gustafsson+Nordebo2013}.

\section{Radiation Modes and Degrees of Freedom}
\label{S:ModesAndDoF}

The maximum gain was expressed by Harrington in spherical mode expansion as~\cite{Harrington1960}
\begin{equation}
  G_\mrm{H} = L^2 + 2L = \frac{N_{\mrm{DoF}}}{2},
\label{eq:maxmodes}
\end{equation}
where~$L$ is the order of the spherical modes and~$N_{\mrm{DoF}}$ degrees of freedom, \ie{}, total number of modes~\cite{Kildal+etal2017}. The maximum gain is related to the size of an antenna aperture~$ka$ by a cut off limit for modes~\mbox{$L = ka$}~\cite{Harrington1961}, but should be corrected for~\mbox{$ka<1$} as~\mbox{$L\geq 1$}, see also~\cite{Kildal+etal2017}. This spherical mode expansion is most suitable for spherical geometries but overestimates the number of modes for other shapes.

In order to take a specific shape of an antenna into account, the modes maximizing the radiated power~$\Prad$ over the lost power~$\Pl$, \ie{}, those minimizing dissipation factor~$\delta$, are found from an eigenvalue problem~\cite{Harrington1960,Jelinek+Capek2017,TEAT-7260} as
\begin{equation}
  \Rmr \Jm_n = \varrho_n \Rs \Psim \Jm_n,
\label{eq:radmodesDef}
\end{equation}
where $\Rml=\Rs\Psim$ was substituted on the right-hand side, and only modes with~\mbox{$\delta_n=\varrho_n^{-1} < 1$} are considered here as well-radiating. It can be seen in~\eqref{eq:radmodesDef} that the eigenvectors~$\Jm_n$ do not change with the surface resistivity and only the eigenvalues have to be rescaled with~$\Rs$. Formula~\eqref{eq:radmodesDef} can be simplified using
\begin{equation}
  \eig(\Rmr,\Psim)
  =\eig(\Sm\Rmlf^{-1}\Rmlf^{-\herm}\Sm^{\herm})  
  =\svd(\Sm\Rmlf^{-1})^2,
\label{eq:radmodes}
\end{equation}
where we also used the factorization~$\Rmr=\Sm^{\herm}\Sm$ based on the spherical mode matrix~$\Sm$,~\cite{Tayli+etal2018}, see Appendix~\ref{S:MoMmatrices}, and a Cholesky factorization~\mbox{$\Psim=\Rmlf^{\herm}\Rmlf$} to reduce the computational burden. The radiation modes in~\eqref{eq:radmodesDef} produce an expansion in modes with orthogonal far fields and increasing dissipation factors. They also appear in the analysis of the eigenvalue problems for the radiation operator~\cite{Schab2016} and for MIMO capacity problems. Notice, that for a spherical shell they form a set of properly scaled spherical harmonics.

\begin{figure}
\includegraphics[width=\columnwidth]{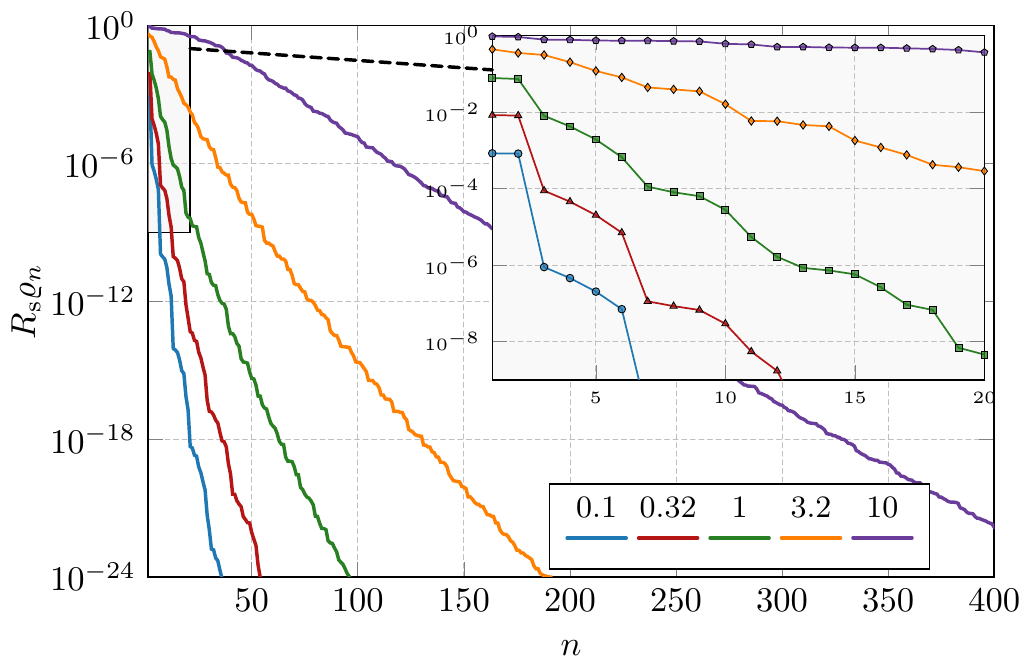}%
\caption{Radiation modes for a rectangular plate of side aspect ratio~$2:1$ and electrical sizes~\mbox{$ka\in \left\{ 0.1,0.32,1,3.2,10 \right\}$}.}%
\label{fig:radmodes}
\end{figure}

Radiation modes~\eqref{eq:radmodesDef} are evaluated for a rectangular plate of side aspect ratio~$2:1$ and electrical sizes \mbox{$ka\in \left\{ 0.1,0.32,1,3.2,10 \right\}$} and the normalized eigenvalues~$\varrho_n \Rs$ are depicted in Fig.~\ref{fig:radmodes}. The low-order modes are emphasized in the inset, where it is seen that the modes appear in groups with similar amplitudes for small~$ka$. This is confirmed via spherical mode expansion, see Appendix~\ref{S:sphere}, for which the rectangular plate supports only half of the spherical modes, \eg{}, $x$- and $y$- electrical and $z$-directed magnetic dipole modes. This characteristic is most emphasized for electrically small structures and the increasing cost of higher order modes vanishes with increasing electrical size, \eg{}, the first ten modes for~\mbox{$ka=3.2$} differ only in a factor of ten compared to~$10^5$ for~$ka=0.32$.

\section{Conclusion}
\label{S:Conclusion}

Maximum gain and effective area for arbitrarily shaped antenna regions are formulated as quadratically constrained quadratic programs (QCQP) which are effectively solved as low-rank eigenvalue problems. The approach is general and includes constraints on self-resonance and parasitic objects, such as reflectors and ground planes. Radiation modes are used to interpret the results and simplify the numerical solution of the optimization problems.

The results are illustrated for a variety of shapes, electrical sizes ranging from subwavelength objects to objects hundreds of wavelengths long, and resistivities covering a wide range from superconductivity to lossy resistive sheets. Plotting the maximal gain versus electrical size reveals three regions. Dipole and Huygens sources dominate in the electrically small region, where the gain depends strongly on the resistivity and whether self-resonance is enforced or not. The effect of self-resonance diminishes as the electrical size approaches a wavelength. End-fire radiation dominates over broadside radiation for objects of wavelength sizes. This changes in the limit of electrically large objects where the effective area is proportional to geometrical cross-section and broadside radiation dominates over endfire radiation. 

Superdirectivity is analyzed from the perspective of determining the trade-off between directivity and efficiency. Here, it is shown that the problem of maximum gain for a given resistivity is solved by the same eigenvalue problem as minimum dissipation factor for a given directivity. Moreover, numerical results suggest that the increase in the dissipation and Q-factor are similar for superdirectivity.     

The results presented in this paper are of general interest as they can be utilized to evaluate the actual performance of designed and manufactured antennas and scatterers with respect to the fundamental bounds. Together with the previously published bounds on Q-factor and radiation efficiency, this work completes the rigorous study of electrically small antenna limits and extends the fundamental bounds towards the electrical large antennas. Understanding of fundamental bounds and knowledge in optimal currents reopen a call for optimal antenna designs.

\appendices

\section{Matrix Representation of Used Operators}
\label{S:MoMmatrices}

The matrices used in the optimization problems are constructed by expansion of the current density~$\Jv(\rv)$ according~\eqref{eq:basFcns} for~$\rv\in\reg$. 

The far-field matrix for direction~$\rvh$ reads~\cite{Gustafsson+etal2016a}
\begin{equation}
 \Fm = \begin{pmatrix}
   \Fm_{\evh} \\
   \Fm_{\hvh} \end{pmatrix},
\label{eq:Fmatrices}
\end{equation}
where $\evh=\hvh\times\rvh$ and $\hvh=\rvh\times\evh$ denote two orthogonal polarizations with elements
\begin{equation}
  F_{\evh,n} = \frac{-\ju k \sqrt{\ZVAC}}{4\pi}\int_{\reg}\evh\cdot\basv_n(\rv_1)\eu^{\ju k\rv_1\cdot\rvh}\diffS_1,
\label{eq:Fmatrix}
\end{equation}
and similarly for~$\Fm_{\hvh}$.

The radiation resistance matrix~$\Rmr$ and reactance matrix~$\Xm$ form the \ac{MoM} \ac{EFIE} impedance matrix~\mbox{$\Zm=\Rmr+\ju\Xm$} of a structure modeled as \ac{PEC} \cite{Harrington1968}.

The ohmic loss matrix~$\Rml = \Rs\Psim$ for a region with a homogeneous surface resistivity, \ie{}, $\Rs$, is given by the Gram matrix~\cite{Horn+Johnson1991}, defined as
\begin{equation}
  \Psi_{mn} = \int_{\reg}\basv_m(\rv)\cdot\basv_n(\rv)\diffS.
\label{eq:grammatrix}
\end{equation}

The expansion matrix between basis functions used and spherical waves reads~\cite{Tayli+etal2018}
\begin{equation}
S_{\sphind n} =  k \sqrt{\ZVAC} \int_\reg \uvop_{\sphind}^{(1)}(k\rv)\cdot \psiv_n(\rv) \diffS\ ,
\label{eq:Smatrix}
\end{equation}
where $\uvop_{\sphind}^{(1)}$ denotes the regular spherical vector waves with index $\sphind$~\cite{Kristensson2016}. The matrix~$\Sm$ is a low-rank factorization of the radiation resistance matrix~$\Rmr=\Sm^{\herm}\Sm$. 

\section{\ac{QCQP}}
\label{S:QCQP}
Maximum gain for self-resonant currents is determined from a \ac{QCQP}~\cite{Boyd+Vandenberghe2004,Park+Boyd2017} of the form~\eqref{eq:maxgain_res}
\begin{equation}
\begin{aligned}
	& \maximize && \Jm^{\herm}\Um\Jm \\
	& \subto &&  \Jm^{\herm}\Rm\Jm = 1 \\
	& && \Jm^{\herm}\Xm\Jm = 0,
\end{aligned}
\label{eq:mG_QCQP}
\end{equation}
where $\Um=\Um^{\herm}\succeq\Om$, $\Rm=\Rm^{\tran}\succeq\Om$, and $\Xm=\Xm^{\tran}$ being indefinite. This formulation can be relaxed to a dual problem 
\begin{equation}
\begin{aligned}
	& \minimize_{\nu}\maximize_{\Jm} && \Jm^{\herm}\Um\Jm, \\
	& \subto &&  \Jm^{\herm}(\nu\Xm+\Rm)\Jm = 1,
\end{aligned}	
\label{eq:mG_QCQPrelax}
\end{equation}
analogously to the analysis in Section~\ref{S:maxGsr} with the solution
\begin{equation}
  \minimize_{\nu} \max \eig(\Um,\nu\Xm+\Rm).
\label{eq:mG_QCQPrelax_sol}
\end{equation}
The range~$\nu\in\left[\nu_\mrm{min},\nu_\mrm{max}\right]$ is restricted such that
\begin{equation}
  \nu\Xm+\Rm\succeq\Om,
\label{eq:PSDregionForNu}
\end{equation}
and an efficient procedure to find~$\nu_\mrm{min}$ and $\nu_\mrm{max}$ is outlined in Appendix~\ref{S:nuregion}.

\begin{figure}
\includegraphics[width=\columnwidth]{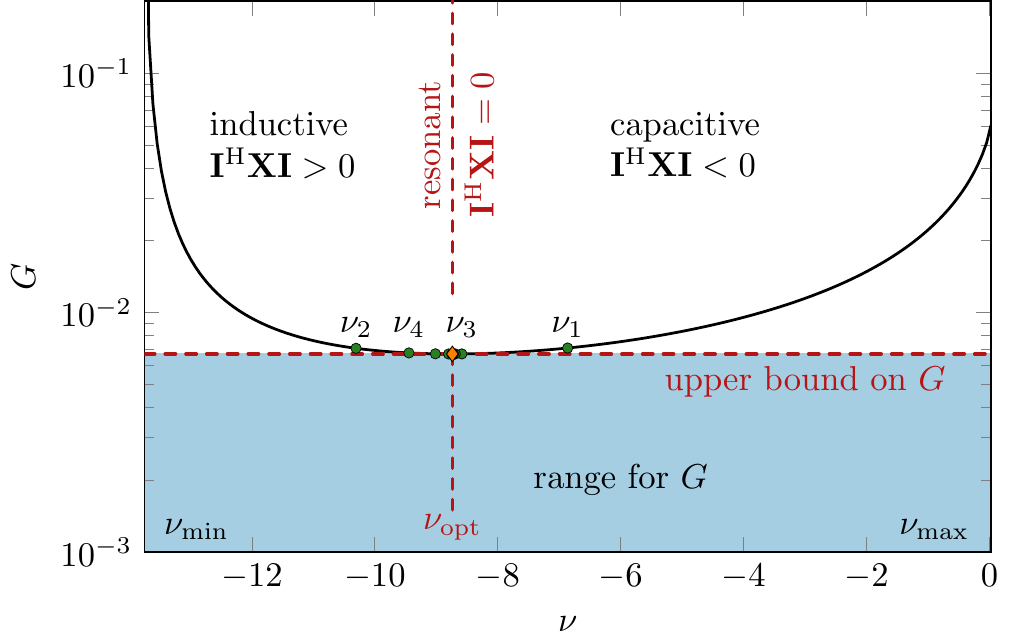}
\caption{Solution of the \ac{QCQP}~\eqref{eq:mG_QCQP} using the dual formulation~\eqref{eq:mG_QCQPrelax_sol}. The range $[\nu_\mrm{min},\nu_\mrm{max}]\approx[-13.7,0.02]$ for the dual parameter~$\nu$ is determined in Appendix~\ref{S:nuregion} by~\eqref{eq:QCQPdualrange} and the optimal parameter value~$\nu_{\mrm{opt}}\approx -8.74$ is determined from the sequence~$\nu_n$, \mbox{$n= \left\{1,2,\ldots \right\}$} using the bisection algorithm~\cite{Nocedal+Wright2006}.}
\label{fig:QCQPGnusweep}
\end{figure}

The minimization problem~\eqref{eq:mG_QCQPrelax_sol} is solved iteratively using a line-search algorithm, \eg{}, the bisection algorithm~\cite{Nocedal+Wright2006}, where also the derivative~\eqref{eq:eigderivative} is used, see Fig.~\ref{fig:QCQPGnusweep} showing the optimization setup. Note that the Newton algorithm~\cite{Boyd+Vandenberghe2004} can be used if the Hessian is evaluated as, \eg{}, in the case with partial gain~\cite{Gustafsson+etal2016a}.

The explicit form of the derivative~\eqref{eq:eigderivative} also shows that the derivative is zero for the optimal value $\nu_{\mrm{opt}}$ if the eigenvalue depends continuously on $\nu$ as the derivative changes sign around $\nu_{\mrm{opt}}$. Hence, the solution to~\eqref{eq:mG_QCQPrelax_sol},~$\Jm_{\mrm{opt}}$, at the extreme point~$\nu_{\mrm{opt}}$ is self resonant $\Jm_{\mrm{opt}}^{\herm}\Xm\Jm_{\mrm{opt}}=0$ and satisfies the second constraint in the \ac{QCQP}~\eqref{eq:mG_QCQP}. This implies that the duality gap is zero and that the \ac{QCQP}~\eqref{eq:mG_QCQP} is solved by its dual~\eqref{eq:mG_QCQPrelax_sol}. Moreover, non-degenerate eigenvalues depend continuously on parameters~\cite{Kato1980} so the problem is solved for this case. For a treatment of modal degeneracies and other implementation issues, see Appendix~\ref{S:QCQPImplementation}.

\section{Alternative Solutions to \ac{QCQP}}
\label{S:QCQPAlternative}

The Lagrangian dual~\cite{Boyd+Vandenberghe2004} is convex and offers an alternative approach to solve the \ac{QCQP}~\eqref{eq:mG_QCQP}. It is given by the \ac{SDP}
\begin{equation}
\begin{aligned}
	& \minimize && \mu, \\
	& \subto &&  
      -\Um+\nu\Xm+\mu\Rm \succeq \Om, \\
  & && \mu\geq 0,\quad \nu\in\R,
\end{aligned}
\label{eq:QCQP_Ldual}
\end{equation}
which can be solved efficiently~\cite{Boyd+Vandenberghe2004}. The semidefinite constraint in the Lagrangian dual~\eqref{eq:QCQP_Ldual} can be written
\begin{equation}
  \Jm^{\herm}(\nu\Xm+\mu\Rm)\Jm
  =\mu\Jm^{\herm}(\nu_1\Xm+\Rm)\Jm
  \geq \Jm^{\herm}\Um\Jm
\label{eq:QCQP_Ldualrange}
\end{equation}
for all currents~$\Jm$ and $\nul=\nu/\mu$. Here, it is seen that $\nul\Xm+\Rm\succeq\Om$ and
\begin{equation}
  \mu \geq \frac{\Jm^{\herm}\Um\Jm}{\Jm^{\herm}(\nul\Xm+\Rm)\Jm} \geq \min_{\nul} \max\eig(\Um,\nul\Xm+\Rm)
\label{eq:LagrangDual}
\end{equation}
and hence the solution of the Lagrangian dual in~\eqref{eq:QCQP_Ldual} is similar to the solution~\eqref{eq:mG_QCQPrelax_sol} of the relaxation~\eqref{eq:mG_QCQPrelax}. The only difference is in the range for $\nul$ that is a subset of~\eqref{eq:QCQPdualrange} due to the $\Jm^{\herm}\Um\Jm$ term in the right-hand side of~\eqref{eq:QCQP_Ldualrange}. However, self-resonant solutions of~\eqref{eq:maxGD_pareto} satisfies~\eqref{eq:QCQP_Ldualrange}. 
Semidefinite relaxation is another standard relaxation technique~\cite{Boyd+Vandenberghe2004} for the \ac{QCQP}~\eqref{eq:mG_QCQP}.

\section{Numerical Evaluation of \ac{QCQP}}
\label{S:QCQPImplementation}

In this paper, the implementation is as follows. We use the eigenvalue problem~\eqref{eq:mG_QCQPrelax} together with the factorization~\mbox{$\Um=\Fm^{\herm}\Fm$} due to its simplicity and computational efficiency. The computational complexity is dominated by the solution of the linear system~\mbox{$(\nu\Xm+\Rm)^{-1}\Fm^{\herm}$} which requires of the order~$N^3$ operations for direct solvers, where~$N$ is the number of basis functions, \cf{}~\eqref{eq:basFcns}. Here, we also note that the additional computational cost of using multiple directions~$\Fm$ is negligible. For electrically large structures we use iterative algorithms to solve the linear system~\cite{Saad1996}. The rectangle in Fig.~\ref{fig:RecAeff} was, \eg{}, solved iteratively using a matrix-free FFT-based formulation~\cite{Chew+etal2001} using \mbox{$N\approx 4.2 \cdot 10^6$} unknowns for~$ka\approx 10^3$. The \ac{FMM} and similar techniques can also be used~\cite{Chew+etal2001} to reduce the computational burden.

Whenever possible, symmetries are used to simplify the solution by separating the eigenvalue problem into orthogonal subspaces which are solved separately and combined analytically~\cite{TEAT-7260}. This reformulates the optimization problems into block diagonal form, where each block corresponds to a subspace. The problem is further simplified for cases where some of the subspaces do not contribute to the radiation intensity in the considered direction as, \eg{}, for radiation in the normal direction~$\kvh=\zvh$ for the rectangle in Fig.~\ref{fig:RecAeff}, where currents with odd inversion symmetry~\mbox{$\Jv(\rv)=-\Jv(-\rv)$} do not contribute and similarly for the cylinder in Fig.~\ref{fig:CylAeff} where azimuthal Fourier indices~$|m|\neq 1$ do not contribute. For these cases, the currents in the non-contributing subspace can only be used to tune the currents into self resonance. Hence, they are quiescent in the externally tuned case~\eqref{eq:gain_res} and determined by the eigenvectors associated with the largest eigenvalues of the eigenvalue problems in~\eqref{eq:nurange}, where the matrices $\Xm,\Rmr$, and $\Rml$ are restricted to the non-contributing subspace. 

Expansion in radiation modes~\eqref{eq:radmodesDef} is also useful in the solution of the maximum gain optimization problem~\eqref{eq:maxgain}, which contains the solution of the linear system $(\Rmr+\Rs\Psim)^{-1}\Fm^{\herm}$ and often is solved for many values of $\Rs$ as in Section~\ref{S:superdir}. Using $\Rmr=\Sm^{\herm}\Sm$ and $\Rml=\Rs\Rmlf^{\herm}\Rmlf$ together with the singular value decomposition (SVD) $\Um\Sigmam\Vm^{\herm}=\Sm\Rmlf^{-1}$ reduces the inversion of the linear system to inversion of a diagonal matrix, \ie, 
\begin{align}
  (\Rmr &+ \Rml)^{-1} = (\Sm^{\herm}\Sm+\Rs\Rmlf^{\herm}\Rmlf)^{-1} \nonumber \\  
  &= \Rmlf^{-1}(\Rmlf^{-\herm}\Sm^{\herm}\Sm\Rmlf^{-1} +\Rs\Id)^{-1}\Rmlf^{-\herm} \nonumber \\
  &= \Rmlf^{-1}(\Vm^{-\herm}\Sigmam^{\herm}\Sigmam\Vm^{\herm} +\Rs\Id)^{-1}\Rmlf^{-\herm} \nonumber \\
  &= \Rmlf^{-1}\Vm(\Sigmam^{\herm}\Sigmam +\Rs\Id)^{-1}\Vm^{\herm}\Rmlf^{-\herm},
\label{eq:radmodes2}
\end{align}
where~$\Id$ denotes the identity matrix. The computational cost of sweeping the maximum gain versus~$\Rs$ is hence traded to computation of the SVD of~$\Sm\Rmlf^{-1}$ that only requires~$N_{\mrm{s}}^2N+N^2$ operations, where~$N_{\mrm{s}}$ denotes the number of spherical modes. Note, that we also use that matrix~$\Psim$ is a sparse matrix with approximately~$3N$ non-zero elements that reduces the computational cost to compute~$\Rmlf$.

\section{Determination of $\nu_\mrm{min}$ and~$\nu_\mrm{max}$}
\label{S:nuregion}

The task here is to find a range of~$\nu$, delimited by~$\nu_\mrm{min}$ and~$\nu_\mrm{max}$, such that \eqref{eq:PSDregionForNu} holds, $\Rm=\Rm^{\tran}\succeq\Om$, and $\Xm=\Xm^{\tran}$ is indefinite. Equivalently, we can state that
\begin{equation}
  \Jm^{\herm}(\nu\Xm+\Rm)\Jm \geq 0 \quad \forall \, \Jm
\label{eq:PSDregion1}
\end{equation}
which can be reformulated to the Rayleigh quotient
\begin{equation}
  \nu\frac{\Jm^{\herm}\Xm\Jm}{\Jm^{\herm}\Rm\Jm} \geq -1 \quad \forall \, \Jm.
\label{eq:PSDregion2}
\end{equation}

\begin{figure}
\centering
\includegraphics[]{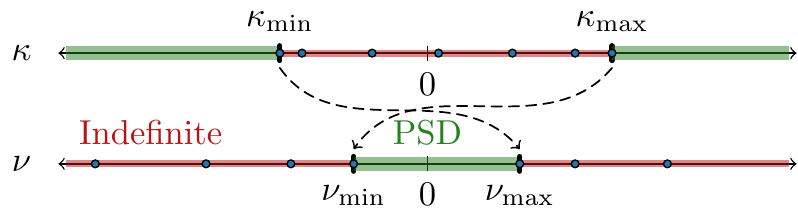}
\caption{Determination of the range of~$\nu: \nu\Xm+\Rm\succeq\Om$; eigenvalues~$\kappa$ are solutions to $\Xm \Jm = \kappa \Rm \Jm$.}
\label{fig:PSDexplanation}
\end{figure}

First, consider the case with $\nu>0$, for which the Rayleigh quotient satisfies
\begin{equation}
  \frac{\Jm^{\herm}\Xm\Jm}{\Jm^{\herm}\Rm\Jm}\geq
  \min_{\Jm}\frac{\Jm^{\herm}\Xm\Jm}{\Jm^{\herm}\Rm\Jm}
  =\min\eig(\Xm,\Rm)
  =\kappa_\mrm{min}
   \geq\frac{-1}{\nu}
\label{eq:PSDpositive1}
\end{equation}
that implies the upper limit of the interval
\begin{equation}
	\nu\leq \nu_\mrm{max}=\frac{-1}{\kappa_\mrm{min}},
\label{eq:PSDpositive2}    
\end{equation}
where~$\kappa_\mrm{min}$ is the smallest eigenvalue, see Fig.~\ref{fig:PSDexplanation}.

Second, consider the case~$\nu<0$. Analogously to~\eqref{eq:PSDpositive1} we get
\begin{equation}
  \frac{\Jm^{\herm}\Xm\Jm}{\Jm^{\herm}\Rm\Jm}\leq
  \max_{\Jm}\frac{\Jm^{\herm}\Xm\Jm}{\Jm^{\herm}\Rm\Jm}
  =\max\eig(\Xm,\Rm)
  =\kappa_\mrm{max}
   =\frac{-1}{\nu}
\label{eq:PSDnegative1}
\end{equation}
and the range is given by \eqref{eq:PSDpositive2} and \eqref{eq:PSDnegative1} as
\begin{equation}
  \frac{-1}{\kappa_\mrm{max}}=\nu_\mrm{min}\leq \nu \leq \nu_\mrm{max}=\frac{-1}{\kappa_\mrm{min}}.
\label{eq:QCQPdualrange}
\end{equation} 

\section{Maximum Gain for a Sphere}
\label{S:sphere}

The optimization problems~\eqref{eq:gain_res} and~\eqref{eq:maxgain_res} are solved analytically for spherical structures using spherical waves~\cite{Kristensson2016}. The matrices in the eigenvalue problems~\eqref{eq:maxgain} and~\eqref{eq:maxgain_res_dual} are diagonalized for spherical modes offering closed-form solutions of the eigenvalue problems. The radiation resistance and ohmic loss matrices have elements~\mbox{$(ka\Rop^{(1)}_{l\tau}(ka))^2$} and $\Rs$, respectively, giving normalized dissipation factors~\mbox{$\delta_{l\tau}/\Rs=(ka\Rop^{(1)}_{l\tau}(ka))^{-2}$}, where $l$~is the order of the spherical mode, $\tau$~the TE or TM type, and $\Rop^{(p)}_{l\tau}$ radial functions~\cite{Hansen1988}. Radiation dominates for modes with $(ka\Rop^{(1)}_{l\tau}(ka))^{2}>\Rs$. This resembles~\eqref{eq:maxmodes} with the observation that the radial functions are negligible for~$ka\ll l$.

\begin{figure}%
\includegraphics[width=\columnwidth]{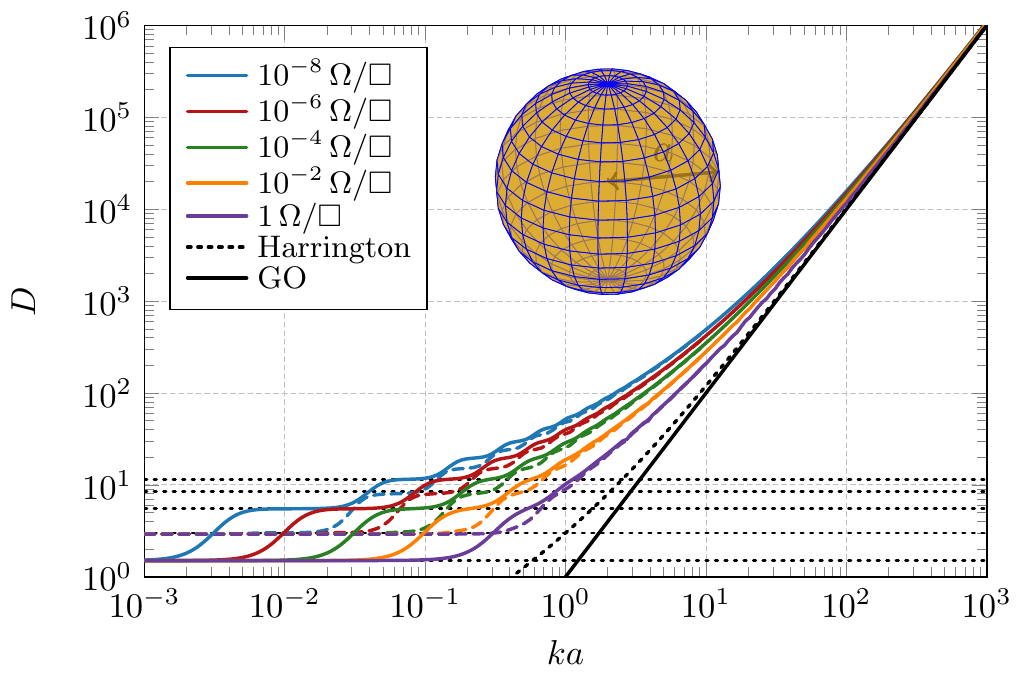}%
\caption{Resulting directivity from the maximum gain for a spherical shell with surface resistivity~$\Rs\in\{10^{-n}\}\Ohmps$ for~\mbox{$n=\left\{0,2,4,6,8\right\}$} in Fig.~\ref{fig:SphereGain}. Dotted lines depicts directivities~$D\in\{3/2,3,11/2,8,23/2\}$ associated with the lowest order modes.}%
\label{fig:SphereGainD}%
\end{figure}

\begin{figure}%
\includegraphics[width=\columnwidth]{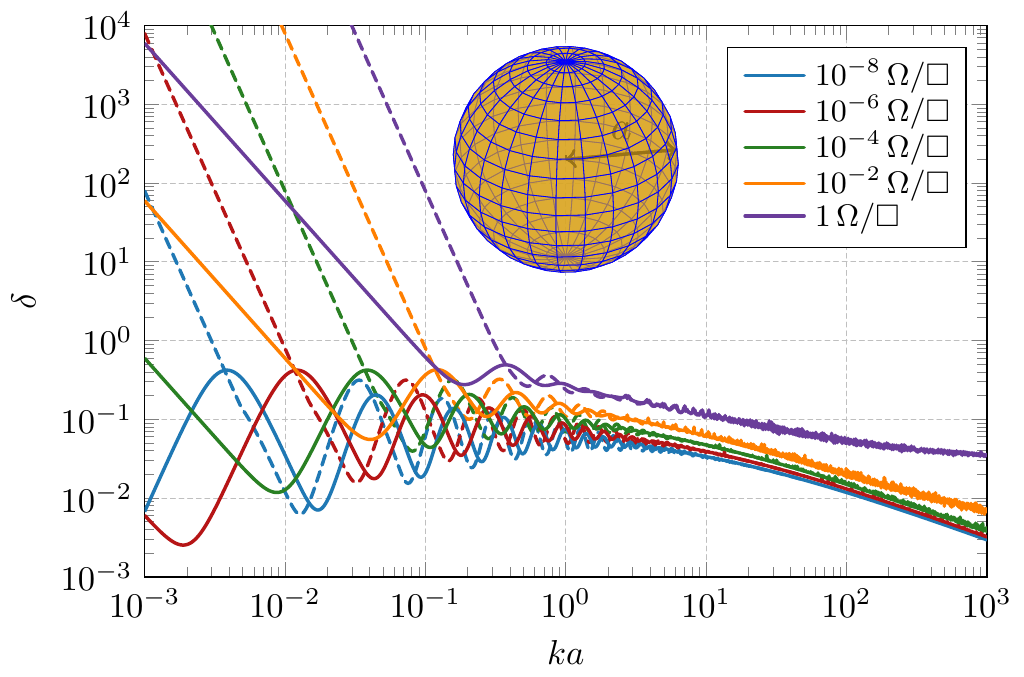}%
\caption{Resulting dissipation factor from the maximum gain for a spherical shell with surface resistivity~\mbox{$\Rs\in\{10^{-n}\}\Ohmps$} for \mbox{$n = \left\{0,2,4,6,8\right\}$} in Fig.~\ref{fig:SphereGain}.}%
\label{fig:SphereGainDelta}%
\end{figure}

The directivity associated with the maximum gain and effective area in Fig.~\ref{fig:SphereGain} is depicted in Fig.~\ref{fig:SphereGainD}. The directivity increases stepwise as additional modes are included. In the electrically small limit, the self-resonant case combines electric and magnetic dipoles to form a Huygens source with directivity~\mbox{$D=3$}. The radiation efficiency is, however, low as seen in the much lower gain in Fig.~\ref{fig:SphereGain}. Inclusion of quadrupole modes increases the directivity to~\mbox{$D=8$} as seen by~\eqref{eq:maxmodes} for~\mbox{$L=2$}. The externally tuned case starts at~\mbox{$D=3/2$}, where the radiation is caused by a sole electric dipole. It increases to~\mbox{$D=11/2$} when the magnetic dipole and electric quadrupole starts to contribute. This stepwise increase is explained by the lower losses for the TM modes, see Fig.~\ref{fig:SphereGainDelta}, causing the modes to appear in order as \mbox{$\{\mrm{TM}_1\}$, $\{\mrm{TM}_2,\mrm{TE}_1\}$, $\{\mrm{TM}_3,\mrm{TE}_2\}$} giving directivity $D=L^2+L-1/2$ as compared with~\eqref{eq:maxmodes}.

\section{Self-Resonant and Tuned Cases}
\label{S:selfResonantTunedCases}
\begin{figure}
\includegraphics[width=\columnwidth]{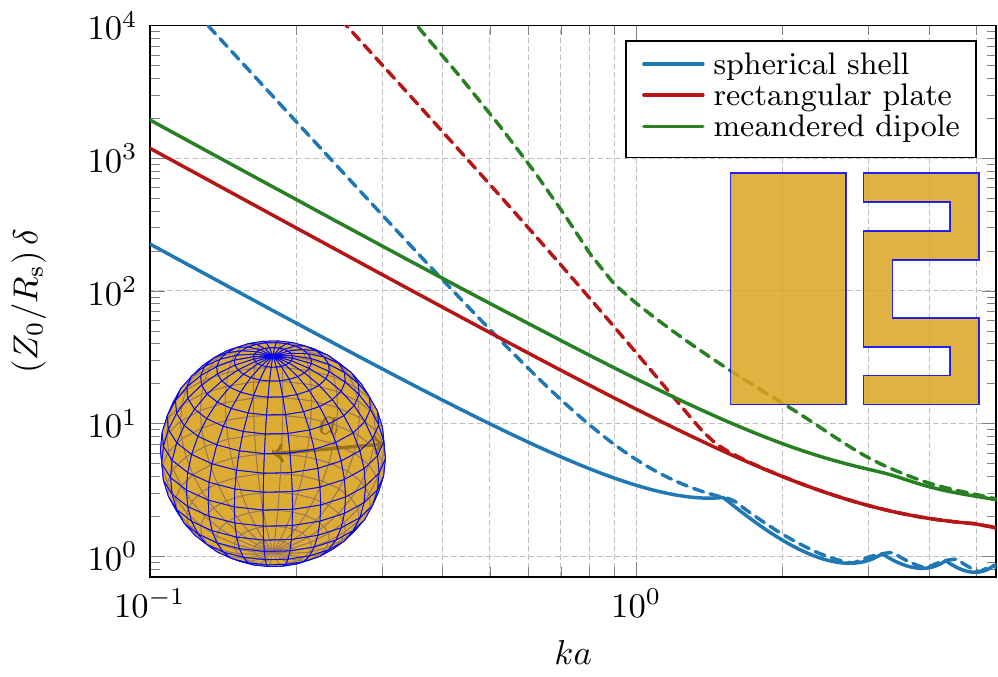}
\caption{Minimum dissipation factors for a spherical shell (blue lines), a rectangular plate (red lines), and a meanderline (green lines) are depicted both for externally tuned (solid lines) and self-resonant (dashed lines) currents.}
\label{fig:MinDFcomp}
\end{figure}

The large difference between the gain (the effective area) for the externally tuned and self-resonant cases for electrically small structures originates in much higher dissipation factor for the loop current forming the TE dipole mode than for the charge separation producing the TM dipole mode~\cite{Pfeiffer2017,Jelinek+Capek2017,Thal2018,TEAT-7260, Jelinek+etal2018}. The difference reduces as the electrical size increases and is negligible for the electrically large structures. The minimum dissipation factor for the two cases can be used as an estimate of the size when the self-resonance condition becomes irrelevant, which is demonstrated in Fig.~\ref{fig:MinDFcomp} for examples of a spherical shell, a rectangular plate, and a meanderline.

\bibliographystyle{IEEEtran}
\bibliography{total}

\begin{biography}[{\includegraphics[width=1in,height=1.25in,clip,keepaspectratio]{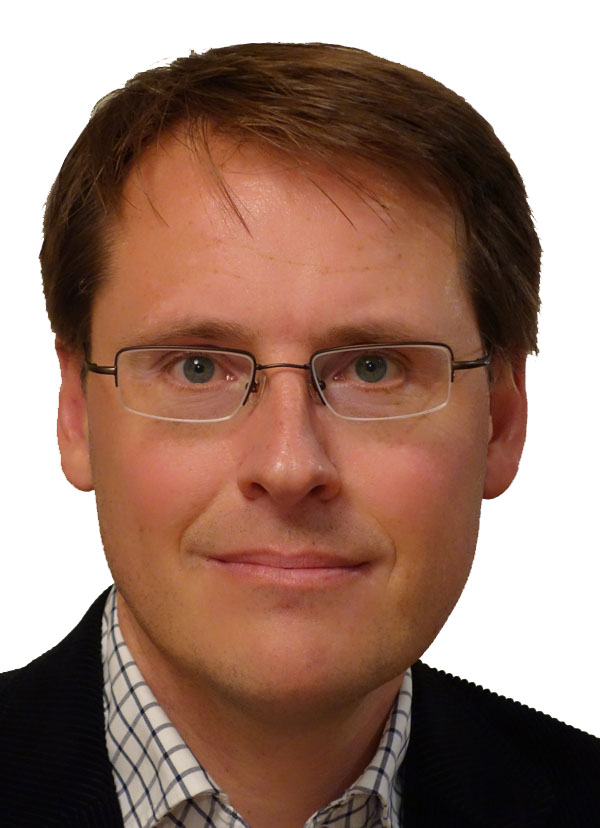}}]{Mats Gustafsson}
received the M.Sc. degree in Engineering Physics 1994, the Ph.D. degree in Electromagnetic
Theory 2000, was appointed Docent 2005, and Professor of Electromagnetic Theory 2011, all
from Lund University, Sweden.

He co-founded the company Phase holographic imaging AB in 2004. His research interests are in scattering and antenna theory and inverse scattering and imaging. He has written over 90 peer reviewed journal papers and over 100 conference papers. Prof. Gustafsson received the IEEE Schelkunoff Transactions Prize Paper Award 2010 and Best Paper Awards at EuCAP 2007 and 2013. He served as an IEEE AP-S Distinguished Lecturer for 2013-15.
\end{biography}

\begin{biography}[{\includegraphics[width=1in,height=1.25in,clip,keepaspectratio]{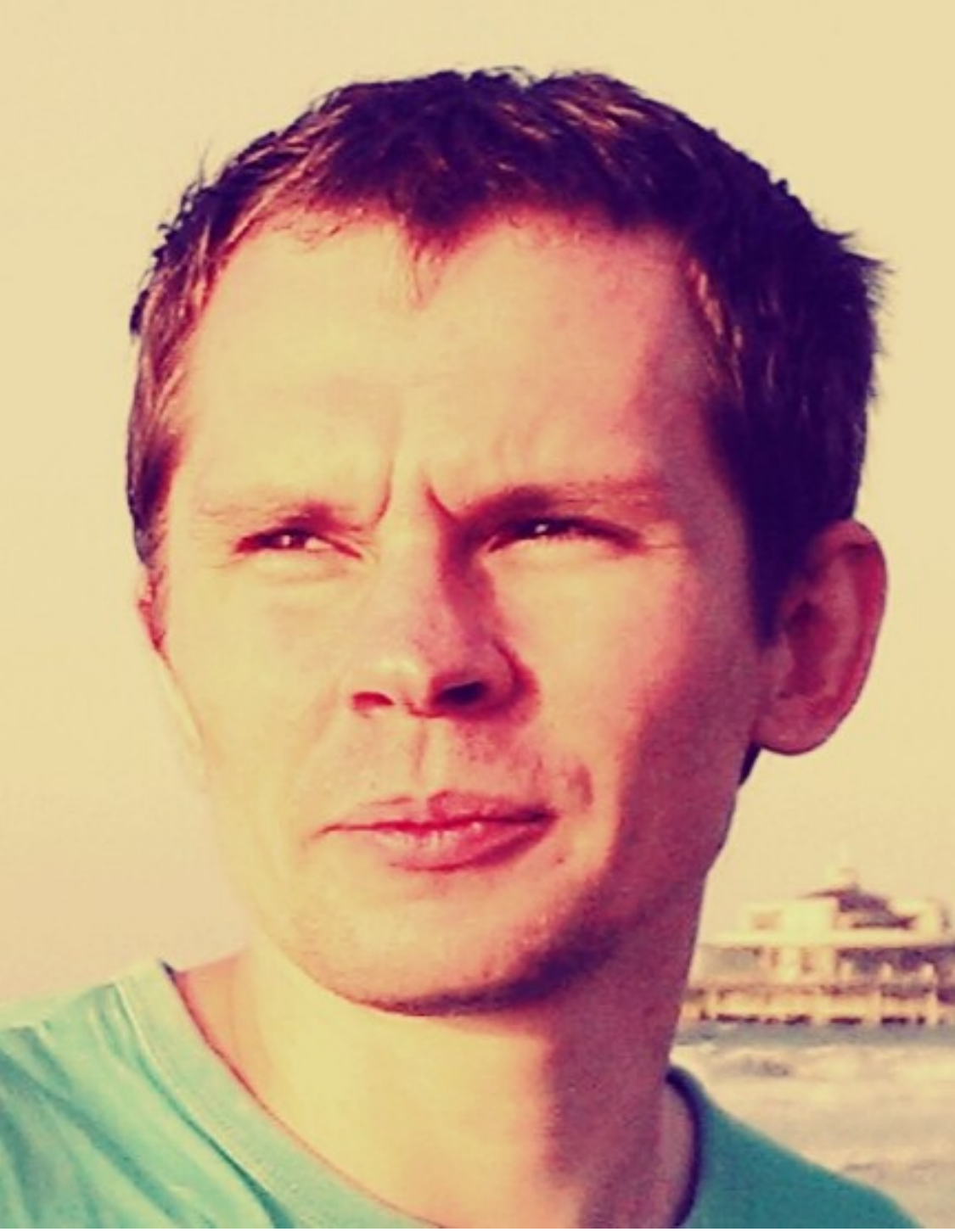}}]{Miloslav Capek}
(M'14, SM'17) received his M.Sc. degree in Electrical Engineering and Ph.D. degree from the Czech Technical University, Czech Republic, in 2009 and 2014, respectively. In 2017 he was appointed Associate Professor at the Department of Electromagnetic Field at the CTU in Prague.
	
He leads the development of the AToM (Antenna Toolbox for Matlab) package. His research interests are in the area of electromagnetic theory, electrically small antennas, numerical techniques, fractal geometry and optimization. He authored or co-authored over 75 journal and conference papers.

Dr. Capek is member of Radioengineering Society, regional delegate of EurAAP, and Associate Editor of Radioengineering.
\end{biography}

\end{document}